\documentclass[12pt]{article}

\usepackage{amsmath}
\usepackage{amssymb}
\usepackage{graphicx}
\usepackage{color}
\usepackage{mathabx}
\usepackage{hyperref}

\usepackage{xcolor}
\usepackage{amsmath,amssymb,epsf,epic}

\usepackage{mathtools}

\usepackage{xr}

\numberwithin{equation}{section}

\newcommand{\ra}{\rightarrow}

\newcommand{\bra}{\langle}
\newcommand{\ket}{\rangle}

\newcommand{\F}{{\mathcal F}}
\renewcommand{\a}{{\rm a}}
\newcommand{\ad}{\ve}
\newcommand{\f}{{\rm f}}
\newcommand{\emin}{{e^{-i\frac{\omega t}{\varepsilon}}}}
\newcommand{\eplu}{{e^{i\frac{\omega t}{\varepsilon}}}}

\newcommand{\G}{{\mathcal G}}

\newcommand{\be}{\begin{equation}}
\newcommand{\ee}{\end{equation}}
\newcommand{\bea}{\begin{eqnarray}}
\newcommand{\eea}{\end{eqnarray}}

\newcounter{resultcounter}[section]

\newtheorem{thm}[resultcounter]{Theorem}
\newtheorem{lem}[resultcounter]{Lemma}
\newtheorem{prop}[resultcounter]{Proposition}
\newtheorem{cor}[resultcounter]{Corollary}

\newtheorem{rem}[resultcounter]{Remark}

\newcommand{\R}{{\mathbb R}}

\newcommand{\C}{{\mathbb C}}

\newcommand{\cx}{{\mathbb C}}
\newcommand{\rx}{{\mathbb R}}
\renewcommand{\i}{{\rm i}}

\def\qed{\hfill $\Box$\medskip}

\newcommand{\bbbone}{\mathchoice {\rm 1\mskip-4mu l} {\rm 1\mskip-4mu l}
{\rm 1\mskip-4.5mu l} {\rm 1\mskip-5mu l}}

\renewcommand{\R}{{\mathcal R}}

\newcommand{\ve}{{\varepsilon}}
\newcommand{\vl}{{\ve,\lambda}}
\renewcommand{\C}{{\mathcal C}}

\begin{document}

\title{The	Adiabatic  Wigner-Weisskopf Model}

\author{ Alain Joye\footnote{ Univ. Grenoble Alpes, CNRS, Institut Fourier, F-38000 Grenoble, France \tt email: alain.joye@univ-grenoble-alpes.fr}\quad  and\quad  Marco Merkli\footnote{Memorial University of Newfoundland, Department of Mathematics and Statistics, St. John's, NL, Canada A1C 5S7, \tt email:\ merkli@mun.ca  } }
\date{October 27, 2022 }
\maketitle

\begin{abstract}
We consider a slowly varying time dependent $d-$level atom interacting with a photon field. Restricted to the single excitation atom-field sector, the model is a time-dependent generalization of the Wigner-Weisskopf model describing spontaneous emission of an atomic excitation into the radiation field. We analyze  the dynamics of the atom and of the radiation field in the adiabatic and small coupling approximations, in various regimes. In particular, starting with an excited atomic state, we provide a description of both the radiative decay of the atom and of the buildup of the photon excitation in the field.
\end{abstract}

\section{Introduction}

This paper is concerned with the dynamics of an open `system-bath' model. The system is a $d$-level `atom', placed in a `bath', or radiation field, which is modeled by a free Bose field. The system Hamiltonian, $H_\a(t)$, depends slowly (adiabatically) on time. Generally, rigorously deriving the effective reduced dynamics of an open quantum system is a primary task in qunatum theory. Even in the easier case when $H_\a$ does not depend on time, a treatment is not simple. In the literature, often the `bath' (radiation field) is considered in a state of thermal equilibrium and one asks whether the coupled system-bath complex converges to the joint, interacting equilibrium in the limit of large times -- a phenomenon called `return to equilibrium'. This effect happens when atomic and field modes are exchanging energy, generating transitions between the system levels leading to the thermal distribution of the original system energies, to lowest order in the system-bath interaction. A different line of inquiry, towards which we aim to contribute here, considers the process of emission of an excitation. In this setting, the interaction between the atom and the field enables transitions between the excited atomic states and its ground state only, not between excited states directly. As a result, the atom is driven, generically, towards its ground state, losing the excitation to the field. It describes the spontaneous emission of an excitation. 

The Wigner-Weisskopf model introduced in \cite{WW} provides a simple description of the exponential in time de-excitation of an atom coupled to a field of photons. It considers a single-excitation process between the levels of the atom and one photon only. The interaction term in the Hamiltonian is a rank two coupling operator, a structure which allows for a mathematically rigorous analysis of the dynamics. Our version of this model includes a possible time dependence (an additional external influence) of the atomic Hamiltonian. We study this generalized Wigner-Weisskopf atom in the adiabatic and small coupling asymptotic regime. Our main results are a detailed expansion of the dynamics of the initially excited atom (the population of each level), as well as the buildup of the excitation in the field in various asymptotic regimes.

From the point of view of adiabatic quantum control, the problem at hand addresses the following situation. Given a smooth time dependent $d$-level Hamiltonian with simple eigenvalues, an initial eigenstate  evolves to a state close to the corresponding instantaneous eigenstate obtained by continuity, provided the time variation of the Hamiltonian is slow enough. This simple version of the adiabatic theorem of quantum mechanics \cite{BF, K1, N1, ASY} allows one to perform quantum engineering on the system provided one has sufficient control on its time dependent Hamiltonian. When the $d$-level system is further subject to interactions with an environment, the adiabatic picture is blurred and a quantification of the effect of the environment becomes of practical interest. This question was addressed in \cite{JMS} for a two-level system coupled to a Bose field by means of an instantaneous energy conserving interaction, in which the interaction operator commutes with the system two-level Hamiltonian at any moment in time. By contrast, the interaction between the $d$-level atom and the field we consider in this present work is much more generic, as it is not assumed to be energy conserving. Accordingly, the dynamics of the atom in the adiabatic and small coupling regime differs from that found in \cite{JMS}, and it is closer to that expected for a generic energy-exchanging model. Indeed, our results are in keeping with those established in \cite{J} for a similar physical situation addressed within the framework of an effective dynamics generated by a Lindbladian with a time dependent Hamiltonian and a generic dissipator, in the corresponding adiabatic and small amplitude of the dissipator regimes. Correspondingly, the results of \cite{JMS} are comparable to those obtained for effective dynamics generated by dephasing Linbdbladians, in the sames asymptotic regimes, see \cite{AFGG, AFGG2}.

\section{Model and main results}

We consider an idealized atom having $d$ possibly time-dependent, simple excited energy levels $0<\alpha_1(t) < \ldots <  \alpha_d(t)$ with normalized eigenstates $\phi_j(t)$ and a ground state energy $\alpha_0=0$ with a corresponding time-independent, normalized eigenstate $\phi_0$. The atom is in contact with a radiation field, modeled by photons with momenta $k\in\rx^3$ in a state of thermal equilibrium at temperature $T\ge 0$. We assume that the atomic energy spectrum has a gap,
\begin{equation}
	\label{gap}
\inf_{t\ge 0} \min_{j,k: \,j\neq k} |\alpha_j(t)-\alpha_k(t)|=\Delta_0>0.
\end{equation}
The interaction allows for excitations to be exchanged between the atom and the field, such that the total number of excitations is conserved. The total (possibly time-dependent) Hamiltonian, acting on the Hilbert space $\cx^{d+1}\otimes\F$, where $\F$ is the Fock space over the single-particle space $L^2(\mathbb R^3,d^3k)$ describing the field, is given by 
\begin{equation}
\label{1}
H(t) = H_\a(t) +H_\f +\lambda V(t),
\end{equation}
where the free atom and field Hamiltonians are
\begin{equation}
H_\a(t) = \sum_{j=1}^{d} \alpha_j(t) |\phi_j(t)\rangle\langle\phi_j(t)|,\qquad H_\f =\int_{\rx^3}\omega(k) a^*(k)a(k)d^3k.
\end{equation}
To simplify the notation, we write $H_\a(t)$ and $H_\f$ instead of $H_\a(t)\otimes \bbbone$  and $\bbbone\otimes H_\f$, and 
we recall
$$
H_\a(t) \phi_0= 0.
$$
The quantity $\lambda$ in \eqref{1} is a (real) interaction constant and 
\begin{equation}
\label{3} 
V(t) = \sum_{j=1}^d v_j(t) |\phi_j(t)\rangle\langle\phi_0|\otimes a(g) + {\rm h.c.\ }
\end{equation}
Here, $v_j(t)$ are smooth complex valued functions of time and 
\begin{equation}
\label{3.01}
a(g) =\int_{\rx^3} \overline{g(k)}\,  a(k)d^3k
\end{equation}
for a form factor $g\in L^2(\rx^3,d^3k)$.  The term $|\phi_j(t)\rangle\langle\phi_0|\otimes a(g)$ in the interaction \eqref{3} describes the process of absorption of an excitation from the field accompanied by a transition from the ground state to the $j$th excited state in the atom. 
\medskip

{\em Remark: Positive temperatures.\ } For concreteness, the field is presented to be described by the Fock space $\mathcal F$ over the single-particle space $L^2(\rx^3,d^3k)$. The vacuum state in $\mathcal F$ is the equilibrium state of the field at zero temperature. However, our analysis carries through without modification to the positive temperature case. Indeed, it suffices to replace the Fock space $\mathcal F$ by a new Fock space $\mathcal F_\beta$, constructed over a new single-particle space $L^2(\rx\times S^2)$. The detail of this procedure in our context are explained in \cite{JMS}. 
\medskip

We will work under the following regularity hypothesis:
\medskip

\begin{itemize}
\item[{}]
{\bf Assumption (A).}
The atomic Hamiltonian $\rx^+\ni t\mapsto H_\a(t)\in M_d(\cx)$ and coupling amplitudes $\rx^+\ni t\mapsto v_j(t)\in \cx $ are $\C^\infty(\rx^+)$, with finite derivatives at $t=0_+$. Here $\rx^+=\{x\geq 0\}$.
\end{itemize}

\noindent
Note that assumption (A) together with the gap condition \eqref{gap}, ensures by perturbation theory (see Lemma \ref{lem1}), that under the condition
\begin{equation}
	\label{lambdasmall}
\frac{4\lambda^2}{\Delta_0}\|v\|^2_\infty \|\gamma\|_{L^1}<1,
\end{equation} 
the eigenvalues $\alpha_j(t)$ of $H_\a(t)$ are simple, smooth, and the normalized eigenvectors $\phi_j(t)$ can be chosen to form a smooth orthonormal basis of $\cx^d$. In \eqref{lambdasmall}, 
\begin{equation}
    \label{25.1}
\gamma(t)\equiv \langle g, e^{-i\omega t}g\rangle_{L^2}=\int_{\rx^3} e^{-i\omega(k)t}|g(k)|^2 d^3k,\qquad t\in\rx,
\end{equation}
is the field correlation function, and we view $v(t)$ as a vector in $\cx^d$, having components $v_j(t)$. We use the notation, for any time-dependent vector $x(t)\in\cx^d$, 
$$
\|x\|_\infty =\sup_{t\ge 0}\|x(t)\|_{\cx^d}.
$$

\bigskip

{\bf Adiabatic scaling.\ } We consider the time-dependence of the Hamiltonian to be governed by a parameter $\varepsilon>0$, that is, we examine the Schr\"odinger equation
\begin{equation}
\label{15}
i\partial_s \psi_\varepsilon(s) = H(\varepsilon s)\psi_\varepsilon(s), \quad \psi_\ve(0)=\psi^{(0)}, \quad s\in\rx^+.
\end{equation} 
For $\varepsilon$ small, the Hamiltonian varies very slowly in time and undergoes a change of order one on a time scale of order $1/\varepsilon$. 
Introducing the rescaled time $t=\varepsilon s$, the corresponding transformed wave function $\phi(t,\varepsilon)\equiv \psi_\varepsilon(t/\varepsilon)$ satisfies 
\begin{equation}\label{psiad}
i\partial_t \phi(t,\varepsilon) = \frac{i}{\varepsilon}(\partial_t \psi_\varepsilon)(t/\varepsilon) = \frac{1}{\varepsilon}H(t)\psi_\ve(t/\varepsilon) =\frac{1}{\varepsilon}H(t)\phi(t,\varepsilon).
\end{equation}
We thus consider the adiabatic Schr\"odinger equation (rename the wave function $\phi(\cdot,\varepsilon)\rightarrow \psi(\cdot)$)
\begin{equation}
\label{16}
i\varepsilon \partial_t \psi(t) = H(t)\psi(t),  \quad \psi(0)=\psi^{(0)},  \quad t\in\rx^+. 
\end{equation}
If $\psi(t)$ solves \eqref{16} then $\psi(\varepsilon s)$ solves the original \eqref{15}. In the equation involving the rescaled time $t$, \eqref{16}, the Hamiltonian undergoes a change of order one on a time scale of order one.  We will analyze the solutions of \eqref{16}. 

Note that for a time-independent Hamiltonian $H(t)=H$ for all $t$, the solutions of \eqref{15} and \eqref{16}  read
\begin{align}\label{autonom}
\psi_\varepsilon(s)=e^{-isH}\psi^{(0)}, \quad \psi(t)=e^{-\frac{i}{\ve} t H}\psi^{(0)}.
\end{align}
Therefore, in this autonomous setup, we can recover the long time behaviour of the system described by \eqref{16} by setting $t=1$ and considering $\ve \ra 0$, or setting $\ve=1$ and allowing $t\ra\infty$.

In what follows, we will always consider $0<\ve \leq 1$.

\bigskip

{\bf Single excitation manifold. }  The Schr\"odinger evolution \eqref{16} leaves invariant the space of  {\em single excitation} atom-field wave functions
\begin{equation}
{\mathcal P}_1 = \Big\{\psi\in\cx^2\otimes\F\, : \, \psi = \sum_{j=1}^d z_j \, \phi_j(0)\otimes\Omega_\f  + \phi_0\otimes a^*(f)\Omega_\f,\ z_j\in\cx,  f\in L^2(\rx^3,d^3k) \Big\},
\end{equation}
where $\Omega_\f$ is the vacuum state in $\F$. Then the solution of   \eqref{16} within the subspace ${\mathcal P}_1$ has the form
\begin{equation}
\label{sol}
\psi(t) = \sum_{j=1}^d z_j(t) \, \phi_j(0)\otimes\Omega_\f  + \phi_0\otimes a^*(f_t)\Omega_\f
\end{equation}
and since $H(t)$ is self-adjoint, the norm of $\psi(t)$ is conserved,
\begin{equation}
\label{13}
\|\psi(t)\|^2 = \sum_{j=1}^d |z_j(t)|^2 +\|f_t\|^2_{L^2(\rx^3, d^3k)}=1,\qquad t\ge 0.
\end{equation}
We collect the components $z_j(t)$ into a vector, written in the eigenbasis $\{\phi_j(0)\}_{j=1}^d$ of $H_\a(0)$, as
\begin{equation}
z(t) = \begin{pmatrix}z_1(t)\\ \vdots \\ z_d(t)\end{pmatrix} = \sum_{j=1}^d z_j(t)\phi_j(0).
\label{120'}
\end{equation}

At time $t=0$, the probability of finding the system in its excited state $\phi_j(t)$ is simply $p_j(0)= |z_j(0)|^2$. 
The probability of finding the system in the $j$th instantaneous excited state $\phi_j(t)$ at time $t$ is 
\begin{equation}
p_j(t) = \langle \psi(t), \big(|\phi_j(t)\rangle\langle \phi_j(t)| \otimes\bbbone_\f\big) \psi(t)\rangle 
=\Big|\sum_{\ell=1}^d z_\ell(t) \langle \phi_j(t),\phi_\ell(0)\rangle \Big|^2 = \big| \langle \phi_j(t),z(t)\rangle\big|^2.
    \label{121}
\end{equation}
The sum $\sum_{j=1}^np_j(t)$ is the probability to find the excitation in the atom (that is, in any of the instantaneous excited states) at time $t$, and one gets from \eqref{121}, and since $\{\phi_j(t)\}_{j=1}^d$ is an orthonormal basis,
\begin{equation}
\sum_{j=1}^d p_j(t) = \sum_{j=1}^d |z_j(t)|^2 =\|z(t)\|^2.
\label{122}
\end{equation}

We denote both inner products of $\cx^d$ and of $L^2(\rx^3, d^3k)$ by $\langle \cdot, \cdot\rangle$, the arguments making it clear which space is meant.

\subsection{Main results}

We present our main results below in this section. We discuss them in detail and compare them to previous results in Sections \ref{sect:discussion} and \ref{prelit}. 

The field correlation function, defined by \eqref{25.1} and its Fourier transform,
$$
\widehat\gamma(\alpha) = \frac{1}{\sqrt{2\pi}} \int_{\rx}e^{i\alpha t}\gamma(t)dt,
$$
play an important role in the dynamics. We will assume the following decay and regularity hypotheses on $\gamma$.

\begin{itemize}
\item[{}]
{\bf Assumption (B).}
The field correlation function $\gamma(t)$ \eqref{25.1} belongs to $L^1(\rx)$, and further satisfies 

i) $t\mapsto t^2\gamma(t)\in L^1(\rx^+)$, $t\mapsto \partial_t\gamma(t)\in L^1(\rx^+)$,

ii) $|\gamma(t)|\leq C_\gamma/(1+t)^m$  $\forall t\in\rx^+$, for some  $m>2$ and some $0<C_\gamma<\infty$.
\end{itemize}

\noindent

{\em Remark: Positive temperatures.\ } In the positive temperature setting (see the remark after \eqref{3.01}), $\gamma(t)$ is the thermal field correlation function, see for instance \cite{JMS,M1,M2,M3}.

\medskip

Since $\gamma(t)$ is a positive definite, continuous function of $t\in\rx$, Bochner's theorem asserts that the inverse Fourier transform $\widecheck\gamma(\alpha)=(2\pi)^{-1/2}\int_\rx e^{-i\alpha t}\gamma(t)dt\ge 0$ is a positive function. But $\widecheck\gamma(-\alpha) =\widehat\gamma(\alpha)$, so we have 
$$
\widehat\gamma(\alpha)\ge 0,\quad \alpha\in\rx.
$$ 
For $t\ge 0$, $j=1,\ldots,d$, we define the quantities
\begin{equation}
\beta_j(t) = \sqrt{\pi/2}\, |v_j(t)|^2\, \widehat \gamma\big(\alpha_j(t)\big)\ge 0\quad \mbox{and}\quad \widetilde\alpha_j(t) = \sqrt{2\pi} |v_j(t)|^2\, {\rm Im} \widehat{(\chi_+\gamma)}\big(\alpha_j(t)\big),
\label{177}
\end{equation}
where $\chi_+(t)$ is the indicator function of $[0,\infty)$,  as well as the {\em Berry phase} (Lemma \ref{berryphase})
\begin{equation}
\label{bphase}
\xi_j(t) = i\int_0^t \langle \phi_j(u)|\partial_t\phi_j(u)\rangle du\in\rx.
\end{equation}

Our first result gives an expansion of the vector $z(t)$, \eqref{120'}, when the excitation is initially entirely concentrated in the atom, meaning that $\|z(0)\|=1$ (c.f. \eqref{122}).

\begin{thm}[Dynamics of the atom]
\label{leadingorder} Assume {\em (A)} and {\em (B)},
take $z(0)\in\cx^d$ with $\|z(0)\|=1$ and suppose that \eqref{lambdasmall} holds. Then, for all $0<\ve\leq 1$,
\begin{equation}
\sup_{0\le t\le 1} \big\| z(t) -  \sum_{j=1}^d  e^{-\frac i\ve \int_0^t [\alpha_j(u)+\lambda^2\widetilde\alpha_j(u) ]du}e^{-\frac{\lambda^2}{\ve} \int_0^t \beta_j(u)du}\, e^{i\xi_j(t)} z_j(0)\phi_j(t)  \big\|  \le C \Big( \ve + \lambda^2+ \frac{\lambda^4}{\ve}\Big),
\label{169}
\end{equation}
for a constant $C$ independent of $\ve,\lambda$.
\end{thm}

The approximating dynamics of $z(t)$ contains an  oscillatory phase and a decaying part ($\int_0^t\beta_j(u)du>0$). It describes the decay of the excitation away from the atom into the field. In particular, Theorem \ref{leadingorder} implies the following estimate on the probabilities of the instantaneous excited levels,    $p_j(t)=|\bra \phi_j(t)|z(t)\ket|^2$ (see \eqref{121}.

\begin{cor}[Population of excited atomic levels]
\label{cor2.2}
Under the conditions of Theorem \ref{leadingorder}, the probability $p_j(t)$ of finding the atom in the  instantaneous excited state $\phi_j(t)$ at time $t$, for $j=1,\ldots,d$, satisfies
\begin{equation}
p_j(t) =  e^{-2\frac{\lambda^2}{\ve} \int_0^t \beta_j(u)du}\, p_j(0)\, +O( \ve + \lambda^2+ \frac{\lambda^4}{\ve}).
\label{175} 
\end{equation}
\end{cor}

Our derivation of the above results for time-dependent $H_a(t)$, can also be used to analyze the case when $H_\a$ is constant in $t$.  The expressions \eqref{177} become time-independent, and we set
\begin{equation}
\label{aaa'}
\alpha'_j = \widetilde\alpha_j-i\beta_j = \sqrt{2\pi} |v_j|^2 {\rm Im}\widehat{(\chi_+\gamma)}(\alpha_j)-i \sqrt{\pi/2} |v_j|^2  \widehat\gamma(\alpha_j).
\end{equation}
Let $P_j$ be the spectral projection of $H_a$ onto the eigenvalue $\alpha_j$ of $H_\a$. 

\begin{cor}[Atom evolution for time-independent $H_a$] 
\label{cor1.13}
Assume {\em (B) ii)} and take $z(0)\in\cx^d$ with $\|z(0)\|=1$. There is a $\lambda_0>0$ such that for $\lambda\leq  \lambda_0$,
\begin{equation*}
\sup_{t\ge 0}\big\| z(t)- \sum_{j=1}^d e^{-it(\alpha_j+\lambda^2 \alpha'_j)} P_j z(0)\big\| \leq 
C\lambda^2,
\end{equation*}
for a constant $C$ independent of $\lambda$.
\end{cor}

The last Corollary shows that $z(t)$ is approximated by the semigroup $e^{t L}$ on $\cx^d$, generated by the operator $L=-i\sum_{j=1}^d(\alpha_j+\lambda^2\alpha'_j)P_j$, uniformly in time $t\ge 0$. The generator is dissipative, ${\rm Re}L\le 0$. Under the `{\em Fermi Golden Rule condition}' $\min_{1\le j\le d} |v_j|^2\widehat \gamma(\alpha_j)>0$, $e^{tL}$ converges to zero exponentially quickly in $t$, at a rate $\propto\lambda^2$. 
\medskip

In our next result, we analyze the properties of the excitation emitted from the atom into the field. The probability density for the field excitation to have momentum $k\in\rx^3$ at time $t$ is given by $|f_t(k)|^2$, see Section \ref{sect:emitted}. This motivates the analysis of averages of the form $\int_{\rx^3}B(k)|f_t(k)|^2 d^3k$ for suitable test functions $B(k)$. We consider the class  $B:\rx^3\rightarrow\cx$ such that  
\begin{equation}
\label{testfunctions}
\gamma_B(t) := \int_{\rx^3} B(k)|g(k)|^2e^{-it\omega(k)}d^3k\ \in\ L^1(\rx,dt)
\end{equation}
(compare with \eqref{25.1}). The following is a well-coupledness condition,
\begin{equation}
\inf_{t\ge 0}\beta_j(t)>0,
\label{wellc}
\end{equation}
where we recall that $\beta_j$ is defined in \eqref{177}. Here is our result on the emitted field excitation momentum density distribution $|f_t(k)|^2$:

\begin{thm}[Momentum distribution of emitted excitation]
\label{thm1.9} Assume {\em (A)}, {\em (B)}, and
suppose that the initial excitation is localized entirely on the excited atomic level $j$ ($1\le j\le d$), meaning that $|z_j(0)|=1$ and $z_k(0)=0$ for $k\neq j$.
Suppose also that \eqref{wellc} holds and let $B$ be any test function satisfying \eqref{testfunctions}. Then
\begin{itemize}
\item[\rm (A)]  In the limit $\ve,\lambda, \frac{\lambda^3}{\ve}\rightarrow 0$, $\frac{\lambda^2}{\ve}\rightarrow \infty$,
\begin{equation}
\lim \int_{\rx^3} B(k) |f_t(k)|^2d^3k =   \frac{\widehat\gamma_B(\alpha_j(0))}{\widehat\gamma(\alpha_j(0))}.
\label{limit}
\end{equation}

\item[\rm (B)] In the limit $\ve,\lambda\rightarrow 0$ such that $\frac{\lambda^2}{\ve}=r>0$ is fixed, 
\begin{equation}
\lim \int_{\rx^3} B(k) |f_t(k)|^2d^3k =  \sqrt{2\pi}\, r\,  \int_0^t |\langle w(s),\phi_j(s)\rangle|^2  e^{-2r\int_0^s\beta_j(u)du}\ \widehat\gamma_B\big(\alpha_j(s)\big)\,ds.
\label{limit'}
\end{equation}
\end{itemize}
\end{thm}

\subsection{Discussion of main results}
\label{sect:discussion}

Generally, our results show the details of the emission of the excitation from the atom to the field, meaning that $|z(t)|\rightarrow 0$ as $t\rightarrow\infty$. In the time-independent case ($H(t)=H$), Corollary \ref{cor1.13} shows that the population $p_j(t)=|z_j(t)|^2$ of the excited level $j$ satisfies
$$
p_j(t)= e^{- 2\lambda^2\beta_j t }p_j(0) +O(\lambda^2),
$$
uniformly in $t\ge 0$, and where $\beta_j$ is defined in \eqref{aaa'}. When $H_\a(t)$ depends on time and varies slowly  with adiabatic parameter $0<\varepsilon\le 1$, the decay rates $\beta_j(t)$ become time-dependent (see \eqref{177}) and the population of the atomic excitations follow the estimate \eqref{175} of Corollary \ref{cor2.2}, 
$$
p_j(t) =  e^{-2\frac{\lambda^2}{\ve} \int_0^t \beta_j(u)du}\, p_j(0)\, +O( \ve + \lambda^2+ \frac{\lambda^4}{\ve}).
$$
In neither case will initially unpopulated atomic  excited levels become populated at later times, beyond an amount bounded by the error terms.

\subsubsection{\bf Discussion of Theorem \ref{leadingorder}} For the error term in \eqref{169} to be small both the coupling constant $\lambda$ and the adiabatic parameter $\ve$ {\em as well as the ratio} $\lambda^4/\ve$ have to be small.  We identify {\bf three regimes}:
\begin{itemize}
    \item[1)] {\em Strong coupling  relative to adiabatic speed: $\ve \ll \lambda^2 \ll \sqrt{\ve} \ll 1$. } In this case, the adiabatic time scale of the Hamiltonian is large with respect to the relaxation time of the system induced by field  ($\propto \lambda^{-2}$) and the coupled system has time enough for the field to absorb the initial excitation on the atom. Namely, for $0\le t\le 1$,
$$
\| z(t)\| \leq  e^{- \frac{\lambda^2}{\ve}\min_{1\le j\le d}\int_0^t\beta_j(u)du}+ O(\lambda^4/\ve).
$$
At the `end of the process', when $t=1$ (which corresponds to the physical time $s=1/\ve$ which is very large), the excitation survival amplitude is bounded above by $e^{-\frac{\lambda^2}{\ve}\tau}+O(\lambda^4/\ve)$, which is very small if the coupling is effective, meaning that  $\tau=\sqrt{\pi/2}\min_{1\leq j\leq d}\int_0^1|v_j(u)|^2\hat\gamma(\alpha_j(u))du>0$.

\item[2)] {\em Comparable coupling strength and adiabatic speed: $\lambda^2=\ve\ll 1$. } The estimate \eqref{169} gives 
$$ 
z(t)=\sum_{j=1}^de^{-\int_0^t\beta_j(u) du} e^{-i \int_0^t [\frac 1\ve \alpha_j(u)+ \widetilde\alpha_j(u)]du} e^{i\xi_j(t)}\, z_j(0)\, \phi_j(t)  +O(\ve).
$$
In this regime, in which the coupling constant multiplied by the time scale is fixed, the effect of the field is to decrease the initial amplitude of the excited state of the atom by an explicit finite correction along each level of the atom. This is the analog of the weak coupling regime which Davies considered for the stationary case ($H$ independent of time), \cite{Davies}.

\item[3)] {\em Weak coupling relative to adiabatic speed: $\lambda^2\ll \ve\ll 1$. } We distinguish two subregimes. For $\ve^2\ll \lambda^2\ll \ve \ll 1$ an expansion of the exponentials in \eqref{169} gives
\begin{multline*}
z(t)=   \sum_{j=1}^d e^{-\frac i\ve \int_0^t \alpha_j(u)du} e^{i\xi_j(t)} z_j(0) \phi_j(t)\\ \times\Big(1-i\frac{\lambda^2}{\ve} \int_0^t \widetilde \alpha_j(u)du- \frac{\lambda^2}{\ve}\int_0^t\beta_j(u) du\Big)  + O\big(\ve+\frac{\lambda^4}{\ve^2}\big),
\end{multline*}
where the integral terms are significant relative to the remainder (as $\ve^2\ll\lambda^2$). In this regime, we find an explicit decrease of the initial amplitude of the excitation of the atom of the order $\lambda^2/\ve$. When this ratio tends to zero the initial excitation is not transferred to the field, the atom stays excited, $\|z(t)\|\sim 1$.

In the second regime, $\lambda^2 \le \ve^2\ll \ve\ll 1$, the expansion of the exponentials in \eqref{169} gives 
$$  
z(t)=   \sum_{j=1}^d e^{-\frac i\ve \int_0^t \alpha_j(u)du} e^{i\xi_j(t)}\, z_j(0) \phi_j(t) +O(\ve).
$$
Again, $\|z(t)\|\sim 1$ and the atom does not get de-excited. In both regimes, the interaction with the reservoir is too weak to significantly alter the dynamics of the atom alone, which evolves adiabatically with its own Hamiltonian. 
\end{itemize}

Recall that we consider initially excited atoms, meaning that $\|z(0)\|^2=1$. The probability for the atom to be in the ground state $\phi_0$ at time $t$ is given by $1-\|z(t)\|^2$, see \eqref{121}, \eqref{122}. We thus define the de-excitation probability at time $t$ as
\begin{equation}
p_\downarrow(t) =1-\|z(t)\|^2.
\end{equation}
Our findings in the different regimes 1)-3) discussed above then imply the following.
\begin{itemize}
 \item[1')] For $\ve \ll \lambda^2 \ll \sqrt{\ve} \ll 1$ we have virtually full de-excitation: $p_\downarrow(t)\ge 1-C\frac{\lambda^8}{\ve^2}\sim 1$.
 
 \item[2')] For $\lambda^2=\ve\ll 1$ we have partial de-excitation: $p_\downarrow(t) = 1-\sum_{j=1}^d   e^{-2\int_0^t\beta_j(u)du }|z_j(0)|^2+O(\ve)$. 
 
 \item[3')] For $\lambda^2\ll \ve\ll 1$ we have virtually no de-excitation:\\
If $\ve^2\ll \lambda^2\ll \ve \ll 1$ then  $p_\downarrow(t)=2\frac{\lambda^2}{\ve} \sum_{j=1}^d |z_j(0)|^2 \int_0^t \beta_j(u)du +O(\ve+\frac{\lambda^4}{\ve^2})\ll 1$.\\
If $\lambda^2 \le \ve^2\ll \ve\ll 1$ then $p_\downarrow(t) = O(\ve)\ll 1$.
\end{itemize}

\medskip

In \cite{JMS} the authors studied the adiabatic transition (scale $\ve$) probability of a time dependent two-level system interacting with a Bose field by means of an instantaneous energy conserving interaction, with coupling constant $\lambda$. 
This kind of coupling induces transitions between the two levels that are likely to be small, since in a time independent situation it conserves  the populations exactly. By contrast, the interaction we consider in the present work allows for instantaneous energy exchange and is more likely to induce transitions between excited and ground states. 

The transition probability from the excited to the ground state for the energy conserving model, found in \cite{JMS}, is given by
\begin{equation}
p_\downarrow^{\rm ec}(t) =\left\{\begin{matrix*}[l]
 \lambda^2\ve \displaystyle\int_0^1 K(s) \widehat\gamma(\Delta(s))ds+o(\lambda^2\ve)   & \mbox{if }\ \ve \ll \lambda^2 \ll \sqrt\ve \ll 1 \\
 & \\
\ve^2 \displaystyle\Big( \displaystyle \int_0^t K(s) \widehat\gamma(\Delta(s))ds + Q(t)\Big)+ o(\ve^2)\quad  &\mbox{if }\ \lambda^2 =\ve\ll 1 \\
 & \\
\ve^2 Q(t)+ o(\ve^2) & \mbox{if }\ \lambda^2 \ll \ve\ll 1 .
  \end{matrix*}\right.
\label{jmsproba}
\end{equation}
Here $K$ and $Q$ are explicit non negative functions constructed from the Hamiltonian, $\Delta(s)>0$ is the energy gap of the two-level system, $\widehat\gamma$ is the Fourier transform of the correlation function of the field. Comparing \eqref{jmsproba} with $p_\downarrow(t)$ we observe the following: 

(a)  $p_\downarrow^{\rm ec}(t)\ll p_\downarrow(t)$ in all regimes, and 

(b) $p_\downarrow^{\rm ec}(t)$ vanishes in all regimes (as $\ve,\lambda\rightarrow 0$), while  $p_\downarrow(t)$ is of order one unless $\lambda^2\ll \ve\ll 1$.
\medskip

The behaviour of $p_\downarrow(t)$ detailed in 1') -- 3') above is similar to that of the adiabatic transition probabilities between the levels of a time dependent gapped Hamiltonian  evolving according to a Lindblad equation modeling a weak reservoir interaction, with a dissipator of order $g>0$, \cite{J}. In this setup, the adiabatic transition probability $p_\downarrow^{\rm L}(t)$ between the excited state of a two-level system at time zero to its ground state at time $t$ was shown in \cite{J} to be
\begin{equation*}
p_\downarrow^{\rm L}(t)=\left\{\begin{matrix*}[l]
R_{1}(t)+O(g^2/\ve+\ve/g)  & \mbox{if }\ \ve \ll g \ll \sqrt\ve \ll 1 \\
\ \\
R_{2}(t) + O(\ve) &\mbox{if }\ g =\ve\ll 1 \\
\ \\
\displaystyle \frac{g}{\ve}\int_0^tJ(s)ds+O(g+\ve^2 + g^2/\ve^2) & \mbox{if }\ g \ll \ve\ll 1 .
  \end{matrix*}\right.
\end{equation*}
Here, the non negative functions $R_1, R_2$ and $J$ depend on the Lindbladian considered, and $R_1(t)\leq 1$. With the identification $g=\lambda^2$, the leading order asymptotics for $p_\downarrow(t)$ and $p_\downarrow^{\rm L}(t)$ have the same behaviour as functions of the coupling constant $g=\lambda^2$ and the  adiabatic parameter $\ve$: they are of order one, unless $\ g=\lambda^2 \ll \ve\ll 1$ in which case they are of order $g/\ve=\lambda^2/\ve$. The actual values of the time dependent coefficients depend on the details of the models.

\subsubsection{Discussion of Theorem \ref{thm1.9}. }

\begin{itemize}
\item[(1)] The meaning of $\widehat \gamma_B(\alpha)$ and energy conservation. We have
\begin{eqnarray*}
\widehat\gamma_B(\alpha) &=& \frac{1}{\sqrt{2\pi}} \int_\rx e^{i\alpha t} \Big[ \int_{\rx^3} \,  B(k)|g(k)|^2 e^{-it\omega(k)} d^3k\Big]dt\\
&=&\frac{1}{\sqrt{2\pi}} \lim_{R\rightarrow\infty} \int_{\rx^3} B(k)|g(k)|^2 \Big[\int_{-R}^R e^{-i(\omega(k)-\alpha)t}dt\Big] d^3k \\
&=& \sqrt{2/\pi}\lim_{R\rightarrow\infty} \int_{\rx^3} B(k)|g(k)|^2 \frac{\sin\big(R(\omega(k)-\alpha)\big)}{\omega(k)-\alpha}d^3k,
\end{eqnarray*}
Commonly one writes $\lim_{R\rightarrow\infty}  \frac{\sin(R(\omega(k)-\alpha))}{\omega(k)-\alpha} = \pi \delta(\omega(k)-\alpha)$, since $\frac{1}{\pi}\sin(Rx/x)$ is a representation of $\delta(0)$ in one dimension as $R\rightarrow \infty$, and so
$$
\widehat\gamma_B(\alpha) = \sqrt{2\pi} B(k)|g(k)|^2 \delta\big(\omega(k)-\alpha\big). 
$$
{\bf Example.\ } Suppose $\omega(k)=|k|$. Then using spherical coordinates $(\omega,\sigma)\in\rx_+\times S^2$, we have for $\alpha>0$,
$$
\int_{\rx^3} B(k)|g(k)|^2 \frac{\sin\big(R(\omega(k)-\alpha)\big)}{\omega(k)-\alpha}d^3k = \int_0^\infty \omega^2 J(\omega) \frac{\sin(R(\omega-\alpha))}{\omega-\alpha}d\omega \rightarrow \pi\alpha^2 J(\alpha)
$$
in the limit $R\rightarrow\infty$, where $J(\omega)=\int_{S^2} B(\omega,\sigma)|g(\omega,\sigma)|^2 d\sigma$, and provided that $J(\omega)$ is differentiable on $\rx_+$ and its derivative satisfies $J'\in L^1(\rx_+,d\omega)$. This illustrates the usefulness of the notation with the delta function.

\item[(2)] The results \eqref{limit}, \eqref{limit'} are a Fermi Golden Rule, describing a process of transition into continuous spectrum. It shows in particular energy conservation (delta function). In the regime (A), the limit is independent of time as the emission process happens right away: $\|z(t)\|\sim e^{-\lambda^2t/\ve}\sim 0$ for all $t>0$ in the limit considered. The emitted momentum density only depends on the energies of the atomic Hamiltonian at time $t=0$. In contrast, in the regime (B), the emission happens gradually, its amplitude grows in $t$ to its final value and the eigenvalue $\alpha_j(s)$ and eigenvector $\phi_j(s)$ contribute to the amplitude, for all $0\le s\le t$.

\item[(3)] We have (see \eqref{177} for $\beta_j(t)$)
\begin{multline*}
\lim_{r\rightarrow\infty}  \sqrt{2\pi}\, r\,  \int_0^t |\langle w(s),\phi_j(s)\rangle|^2  e^{-2r\int_0^s\beta_j(u)du}\widehat\gamma_B\big(\alpha_j(s)\big)\,ds\\
=
\lim_{r\rightarrow\infty}  \sqrt{2\pi}\,   \int_0^{rt} |\langle w(s/r),\phi_j(s/r)\rangle|^2  e^{-2r\int_0^{s/r}\beta_j(u)du}\  \widehat\gamma_B\big(\alpha_j(s/r)\big)\,ds
 =   \frac{\widehat\gamma_B\big(\alpha_j(0)\big)}{\widehat\gamma(\alpha_j(0))}, 
\end{multline*}
so the result \eqref{limit'} is `continuous' with respect to taking $r\rightarrow\infty$ (see \eqref{limit}), morphing the regime (B) into (A).

\item[(4)] For the constant function $B(k)=1$, we have $\gamma_1=\gamma$, which is defined in \eqref{25.1}. It follows from \eqref{limit} that in the parameter regime (A), $\lim \|f_t\|^2_2=1$, which is the probability of emission of the excitation into the field. In the regime (B), this probability depends on time and is given, according to \eqref{limit'}, by 
$$
\lim \|f_t\|^2_2= \sqrt{2\pi}\, r\,  \int_0^t |\langle w(s),\phi_j(s)\rangle|^2  e^{-2r\int_0^s\beta_j}\widehat\gamma\big(\alpha_j(s)\big)\,ds.
$$
\end{itemize}

\subsection{Links to previous literature}\label{prelit}

The dynamics of the Wigner-Weisskopf model in the time {\em in}dependent setup (time independent $H_\a$) was  investigated in details  by means of spectral methods in \cite{Davies}; see also \cite{JKP} for a more recent account. In the time {\em de}pendent framework (time dependent $H_a(t)$) we consider in the current work, we use a different approach,  similar in spirit to that used for deriving master equations, along the lines of \cite{D2, DS}. For detailed information on master equations and the weak coupling limit, the reader may consult \cite{Tr,M1,M2,M3,AL,DF} and their references. The authors of \cite{DS} address the adiabatic dynamics of a time dependent $d$-level system weakly coupled to a Fermi field, corresponding to the regime $\lambda^2=\ve <\!\!< 1$. They derive the asymptotic system state, for a somewhat different class of interactions than we consider here, and they do not analyze the details of the state (such as the evolution of the populations -- which is one of our goals). Related models, in which the bath effect is incorporated into the system Hamiltonian as an effective time dependent term (so-called `classical noise') is commonly used to describe noise assisted quantum excitation transfer processes, see for instance \cite{Ne} and references therein.
Very recently, time dependent variants of the Wigner-Weisskopf model have been used to investigate properties of non-autonomous Lindblad dynamics, with particular focus on their markovian properties \cite{CHL, CL}. Variants of the Wigner-Weisskopf model are used to describe, more abstractly, the coupling of a small quantum system coupled to a large environment characterized by continuous spectrum, see {\it e.g.} \cite{Ma, DF1, AJPP, DK} and references therein.

There are adiabatic approximation results for instantaneous generators with eigenvalues embedded in continuous spectrum \cite{AE,T}. However, in our situation,  the instantaneous Hamiltonian \eqref{1} restricted to the single excitation sector, has purely absolutely continuous spectrum over the whole time span considered, for small enough non zero coupling strength  \cite{Davies, JKP}.  The above mentioned results and tools are thus not amenable for us. In particular, our model and results differ from those of \cite{CJKN}, where the authors consider a Wigner-Weisskopf type model  with a single  ($d=1$) uncoupled excited energy level $\alpha(t)$, varying in such a way that the eigenvalue of the coupled model moves in and out of the absolutely continuous spectrum. They analyze the adiabatic limit of $p_\downarrow(t)$ in the situations where the instantaneous eigenvalue remains embedded in the spectrum,  or where it becomes a resonance. Related results on adiabatic pair creation processes in the Dirac equation are studied in \cite{N2, PD}.

In our master equation approach, on the other hand, we have an effective time dependent generator with simple eigenvalues which. However, the generator is not self-adjoint and also depends on the adiabatic parameter $\ve$ in a singular way. Non self-adjoint generators are known to be amenable to adiabatic techniques, \cite{NR, J2, AFGG, Sch}, but in the present situation, extra care and detail in the analysis is required to control the propagators,  due to the singular $\ve$-dependence.

\section{Proofs of the main results}

\subsection{Dynamics of the atom and proof of Theorem \ref{leadingorder}}

Let $\{ \phi_j(0)\}_{j=1}^d$ be the fixed, time-independent basis of the excited atomic space $\cx^d$ consisting of normalized excited eigenstates of $H_\a(0)$. With respect to this basis, each instantaneous eigenvector $\phi_j(t)$ is given by a $d$-dimensional time-dependent vector, 
\begin{equation}
\phi_j(t) = \sum_{\ell=1}^d \langle  \phi_\ell(0), \phi_j(t)\rangle \, \phi_\ell(0).
\end{equation}

The atomic Hamiltonian $H_\a(t)$ and the interaction $V(t)$ then take the form 
\begin{align}
H_\a(t) &= \sum_{m,n=1}^d [H_\a(t)]_{m,n}|\phi_m(0)\rangle\langle \phi_n(0)|,\quad [H_\a(t)]_{m,n}=\langle \phi_m(0), H_\a(t) \, \phi_n(0)\rangle,\label{6}\\
V(t) &= \sum_{j=1}^d w_j(t) |\phi_j(0)\rangle\langle\phi_0|\otimes a(g) +{\rm h.c.}, \quad w_j(t) = \sum_{\ell=1}^d v_\ell(t)\langle \phi_j(0), \phi_\ell(t)\rangle.
\label{7}
\end{align}

Note that $v(t)=(v_1(t),\cdots, v_d(t))^T$ in (\ref{3}) is obtained from $w(t)$ by the unitary map (\ref{7}).

The parameters $z_j(t)\in\cx$ and $f_t: k\mapsto f_t(k)\in L^2(\rx^3,d^3k)$ of \eqref{sol} satisfy the following closed system of coupled equations,
\begin{eqnarray}
i\varepsilon\partial_tz_j(t) &=& \sum_{\ell=1}^d [H_\a(t)]_{j\ell}\, z_\ell(t) +\lambda w_j(t) \langle g,f_t  \rangle\label{10}\\
i\varepsilon\partial_t f_t(k) &=&\omega(k) f_t(k)+\lambda \sum_{\ell=1}^d \overline{w_\ell(t)} z_\ell(t) \, g(k).\label{11}
\end{eqnarray}
We write \eqref{10}, \eqref{11} in vector form,
\begin{eqnarray}
i\varepsilon\partial_t z(t) &=& A(t)z(t)+\lambda w(t)\langle g,f_t\rangle,\label{14}\\
i\varepsilon\partial_tf_t &=& \omega f_t +\lambda \langle w(t),z(t)\rangle \, g,
\label{14.1}
\end{eqnarray}
where $A(t)$ is the hermitian $d\times d$ matrix representing the restriction of $H_\a(t)$ to the subspace spanned by $\{\phi_j(0)\}_{j=1}^d$ and 
\begin{equation}
z(t) = \begin{pmatrix}z_1(t)\\ \vdots \\ z_d(t)\end{pmatrix}, \quad  w(t) = \begin{pmatrix}w_1(t)\\ \vdots \\ w_d(t)\end{pmatrix}.
\label{120}
\end{equation}

Let $U_\ad(t,s)$ be the free atomic propagator, solving the evolution equation
\begin{equation}
\label{19}
i\varepsilon \partial_t U_\ad(t,s) = A(t)U_\ad(t,s),\quad U_\ad(s,s)=\bbbone.
\end{equation}
As $A(t)$ is hermitian, $U_\ad(t,s)$ is unitary, $U_\ad(t,s)^*=U_\ad(t,s)^{-1}=U_\varepsilon(s,t)$. We will also use the notation $U_\varepsilon(t)\equiv U_\varepsilon(t,0)$.

For a time-dependent vector $x(t)\in\cx^d$, we set
$$
\|x\|_\infty = \sup_{t\ge 0}\|x(t)\|_{\cx^d}.
$$

\begin{prop}[Expansion of excited state amplitudes]
\label{prop1}
Suppose the initial condition $\psi\in\mathcal P_1$ is such that $f_{t=0}=0$, so the excitation is initially in the atom. Then
	\begin{equation}
	\label{20.1}
	i\varepsilon \partial_tz(t) = \big[A(t) -i\lambda^2 w(t)\langle Q_\varepsilon(t)w(t)| \big] z(t) + R_0(t,\lambda,\varepsilon),
	\end{equation}
	where
	\begin{equation}
	\label{22}
	Q_\varepsilon(t) = \frac{1}{\varepsilon}\int_0^t \overline\gamma\Big(\frac{t-s}{\varepsilon} \Big)U_\ad(t,s)ds
	\end{equation}
	with a remainder satisfying 
\begin{equation}
	\|R_0(t,\lambda,\varepsilon)\| \le \lambda^4\ \|v\|_\infty^4 \|\gamma\|_{L^1}\, \|t\gamma(t)\|_{L^1} +\lambda^2\varepsilon\  \|w\|_\infty\ \|\partial_tw\|_\infty\ \|t\gamma(t)\|_{L^1}.
	\label{R0}
\end{equation}
\end{prop}
\bigskip

In the time-independent case $A(t)=A$ we have $U_\ve(t,s)=e^{-\frac{i}{\ve}(t-s)A}$ and a simple change of variables in \eqref{22} gives $Q_\ve(t) = \int_0^{t/\ve}e^{-i x A}\bar\gamma(x)dx$, see Section \ref{timind}. In the general, time-dependent case $A(t)$, we can expand the operator $Q_\varepsilon(t)$,  \eqref{22} using usual methods of adiabatic dynamics, involving Kato's intertwining operator, at the cost of additional $\varepsilon$-dependent error terms. To do so, we define the operator
\begin{equation}
	\label{1.24}
\Gamma_\ve(t) = \int_0^{t/\ve}e^{i xA(t)}\, \gamma(x)dx.
\end{equation}
Then we have the following result. 
\begin{prop}
	\label{Qlem} 
	We have
\begin{equation}
\label{28.2}
\Big\| Q_\ve(t) - \Gamma_\ve(t)^*\Big\|  \le \ve  C_1 +\ve^2 C_2 
\end{equation}
where
\begin{eqnarray}
C_1 &=& c\frac{d^2}{\Delta_0} \max_j \|\partial_t P_j\|_\infty \|\gamma\|_{L^1}+d^2\max_j\|\partial_tP_j\|_\infty \|t\gamma\|_{L^1}+ d\max_j \|\partial_t\alpha_j\|_\infty\,  \|t^2\gamma\|_{L^1}\nonumber\\
C_2 &=& c\frac{d^2}{\Delta_0}\|t\gamma\|_{L^1} \Big[ (1+d \|A\|_\infty) \max_j\|\partial_tP_j\|^2_\infty 
+ \max_j\|\partial_t^2P_j\|_\infty  \nonumber\\
&& \qquad + \frac{\max_j|\partial_t\alpha_j|_\infty}{\Delta_0} \max_j\|\partial_t P_j\|_\infty
\Big],
\end{eqnarray}
for a numerical constant $c$.
\end{prop}

By `numerical constant $c$' we mean a constant $c>0$ which can be taken as an `absolute' integer, not depending on any of the parameters of the problem, such as $A,d,\ve,t,\lambda$. 

\bigskip

Combining \eqref{28.2} with \eqref{20.1} gives the following evolution equation for $z$,
\begin{equation}
\label{20.2}
i\varepsilon \partial_tz(t) = G_\vl(t) z(t) + R_1(t,\lambda,\varepsilon),
\end{equation}
where
\begin{equation}
	\label{20.2.2}
G_\vl(t) = A(t)-i\lambda^2 |w(t)\rangle \langle w(t)| \, \Gamma_\ve(t),
\end{equation}
where the remainder has the bound (use $\ve^2\le\ve$ in \eqref{28.2})
\begin{equation}
\|R_1(t,\lambda,\varepsilon)\| \le \|R_0(t,\vl)\| +\ve\lambda^2\|w\|^2_\infty (C_1+C_2).
\label{1.29}
\end{equation}
As we are interested in the possible decay of $\|z(t)\|$ that would describe the de-excitation of the atom, we want to study the contraction properties of $G_\vl(t)$, \eqref{20.2.2}. For each $t$ fixed, the Lumer-Phillips theorem says that $G_\vl(t)$ generates a contraction semigroup if and only if ${\rm Re}\, G_\vl(t)\leq 0$ ($G_\vl(t)$ is dissipative), see {\it e.g} \cite{EN}. However, the latter property fails to hold for any non-selfadjoint rank one perturbation of $A(t)=A(t)^*$ that does not commute with $A(t)$. Hence, in order to control the propagator generated by $G_\vl(t)$, we further simplify it by perturbation theory in $\lambda$. 

\begin{lem}[Analytic perturbation theory for $G_\vl(t)$]
	\label{lem1}
Suppose that \eqref{lambdasmall} holds. Then we have the following, for any values of $t\ge 0$, $1\geq \ve> 0$:
\begin{itemize}
\item[\rm (a)] $G_\vl(t)$ has simple eigenvalues $\alpha_j(t,\vl)$, $j=1,\ldots,d$, each one lying close to an eigenvalue $\alpha_j(t)$ of $A(t)$, satisfying 
\begin{equation}
	\label{1.42}
	\big|\alpha_j(t,\vl) - \big(\alpha_j(t)+\lambda^2\alpha'_j(t,\ve)\big)\big|\le 8\frac{d}{\Delta_0^2} \lambda^4\|v\|^4_\infty \|\gamma\|^2_{L^1}(\|A\|_\infty + \Delta_0/2), 
\end{equation}
where
\begin{equation}
	\label{1.43}
	\alpha_j'(t,\ve) = -i |v_j(t)|^2 \int_0^{t/\ve} e^{i x\alpha_j(t)}\gamma(x) dx.
\end{equation}

\item[\rm (b)] The rank-one spectral projections $P_j(t)$ and $P_j(t,\vl)$ of $A(t)$ and $G_\vl(t)$, respectively, for $j=1,\ldots,d$, satisfy
\begin{equation}
		\label{1.30}
\big\| P_j(t,\vl)-P_j(t)\big\|\le 	\frac{4\lambda^2}{\Delta_0}\|v\|^2_\infty \|\gamma\|_{L^1}.
\end{equation}
\item[\rm (c)] We have $\|P_j(t,\vl)\|\le 2$ and the time derivative of this projection  has the bound
\begin{eqnarray}
\|\partial_tP_j(t,\vl)\| &\le& \frac{8}{\Delta_0} \Big[ \|\partial_tA\|_\infty + \lambda^2\big( 2\|\partial_t w\|_\infty\|w\|_\infty \|\gamma\|_{L^1}+\|v\|_\infty^2 \|t\gamma(t)\|_{L^1}\, \|\partial_tA\|_\infty \big)\nonumber\\
&& \qquad + \|w\|_\infty ^2 \frac{\lambda^2}{\ve}  |\gamma(t/\ve)|  \Big].
\label{1.34}
\end{eqnarray}
Moreover, there is a constant $C$ (for which one can give an expression similar to the right side of \eqref{1.34}, see \eqref{2.44-1}), independent of $\vl$, such that
\begin{equation}
\|\partial_t P_j(t,\vl) - \partial_t P_j(t) \|= C\lambda^2\big(1+\frac1\ve|\gamma(t/\ve)|\big)
\label{1.36-1}
\end{equation}
and
\begin{equation}
\|\partial^2_t P_j(t,\vl)\|\le C \Big( 1+\lambda^2 +\frac{\lambda^2}{\ve^2} (|\gamma(t/\ve)|(1+t) +|(\partial_t\gamma)(t/\ve)| +\frac{\lambda^4}{\ve^2}|\gamma(t/\ve)|^2\Big).
\label{1.35.3}
\end{equation}
\end{itemize}
\end{lem}

Using the spectral representation $G_\vl(t) = \sum_{j=1}^d \alpha_j(t,\vl)P_j(t,\vl)$ and \eqref{1.42} we get from \eqref{20.2},
\begin{equation}
	\label{20.2.1}
	i\varepsilon \partial_tz(t) = \G_\vl(t)z(t) + R_2(t,\lambda,\varepsilon),
\end{equation}
where
\begin{equation}
	\label{1.35.2}
\G_\vl(t)= \sum_{j=1}^d \big( \alpha_j(t)+\lambda^2\alpha'_j(t,\ve)\big) P_j(t,\vl)
\end{equation}
and the remainder satisfies the bound (use \eqref{1.42} and $\|P_j(t,\vl)\|\le 2$),
\begin{equation}
\|R_2(t,\vl)\| \le \|R_1(t,\vl)\|+16\frac{d^2}{\Delta_0^2} \lambda^4\|v\|^4_\infty \|\gamma\|^2_{L^1}(\|A\|_\infty + \Delta_0/2). 
\label{1.36.2}
\end{equation}
The right hand side of \eqref{1.35.2} is  the spectral representation of the operator $\G_\vl(t)$. 

\medskip
We refrain from writing down the explicit dependence of all estimates in the various norms of the functions that define the problem. Instead, in the following, all quantities that are independent of $t\geq 0$, $0\leq \ve\leq 1$, $\lambda$ are denoted by the generic symbols $C$, $C_0$, $C_1 \, \dots$, which may vary from line to line.    
\medskip

Our next task is to integrate the equation \eqref{20.2.1}. Denote the  evolution operator associated to the linear part of \eqref{20.2.1} by  $U_{\ve,\lambda}(t,s)$,
\begin{equation}
	\label{35}
	i\ve \partial_t U_\vl(t,s)= \G_\vl(t) U_\vl(t,s), \qquad U_\vl(s,s)=\bbbone.
\end{equation}
As the left side carries a factor $\ve$ in front of the derivative, this equation is amenable to the usual adiabatic treatment, even though here the non-self-adjoint generator $\G_\vl(t)$ depends on $\ve$ in a singular way, see \eqref{20.2.2}, \eqref{1.43}. Denote the associated Kato generator by
\begin{equation}
	\label{1.38.1}
K_\vl(t) = \sum_{j=1}^d \big(\partial_tP_j(t,\vl)\big)P_j(t,\vl)=- \sum_{j=1}^d P_j(t,\vl)\big(\partial_tP_j(t,\vl)\big)
\end{equation}
and denote the Kato intertwining operator by $W_\vl(t,s)$, which is the solution of
\begin{equation}
	\label{38}
\partial_t W_\vl(t,s) = K_\vl(t) W_\vl(t,s),\qquad W_\vl(s,s)=\bbbone.
\end{equation}
The latter satisfies the intertwining relations, see {\it e.g.} \cite{Kr, K2} 
\begin{equation}
W_\vl(t,s)P_j(s,\vl) = P_j(t,\vl)W_\vl(t,s),\qquad j=1,\ldots, d.
\end{equation}

The {\em adiabatic} evolution operator $V_\vl(t,s)$ is the solution of the equation
\begin{equation}
	\label{40}
i\ve\partial_t V_\vl(t,s)= \big[\G_\vl(t) + i\ve K_\vl(t)\big] V_\vl(t,s),\qquad  V_\vl(s,s)=\bbbone.
\end{equation}

\begin{prop}
	\label{prop4} Under the condition \eqref{lambdasmall}, the adiabatic evolution has the following properties.
\begin{itemize}
\item[1.] $V_\vl(t,s)$ has the decomposition
\begin{eqnarray}
 V_\vl(t,s) &=& W_\vl(t,s) \Psi_\vl(t,s),\nonumber\\
 \Psi_\vl(t,s) &=& \sum_{j=1}^d P_j(s,\vl)e^{-\frac i\ve \int_s^t [\alpha_j(u)+\lambda^2\alpha'_j(u,\ve)]du}.
 \label{41}
\end{eqnarray}
\item[2.]  There are constants $C_1, C_2$ such that for any $0<\ve\leq 1$, $0\leq s\leq  t $, 
\begin{equation}\label{estVU}
\| V_\vl(t,s) \|\leq 2d e^{(t-s)C_1+\lambda^2C_2}, \ \ \| U_\vl(t,s) \|\leq 2d\, e^{(t-s)C_1+\lambda^2C_2}.
\end{equation}
\item[3.]  $V_\vl(t,s)$ approximates the dynamics $U_\vl(t,s)$ \eqref{35} as follows. Further assuming that 
\begin{equation}
	|\gamma(x)| \le \frac{C_\gamma}{(1+|x|)^m}, \qquad \mbox{for some $m>2$,}
	\label{1.59.1}
\end{equation}
we have for any $0\le s\le t$
\begin{equation}
	\label{42}
\Big\| V_\vl(t,s)- U_\vl(t,s)\Big\|\le C{e^{\tilde C(t-s)}}(\ve +\lambda^2), 
\end{equation}
for constants $C, \tilde C$ independent of $\vl$. 
\end{itemize}
\end{prop}
\begin{rem}
The construction and the properties of the adiabatic evolution obviously hold for $\lambda=0$ as well, in which case $\G_{\ve, 0}(t)=A(t)=A(t)^*$ is the atomic Hamiltonian and $U_{\ve,0}(t,s)=U_\ve(t,s)$ is the evolution \eqref{19}. The eigenprojectors $P_j(t)$ are orthogonal and yield the $\ve-$independent anti-symmetric Kato generator $K(t)=\sum_{j=1}^d \big(\partial_tP_j(t)\big)P_j(t)$. In turn, the corresponding Kato intertwining operator $W(t,s)$ is unitary and independent of $\ve$, while the unitary phase operator reads $\Psi_\ve(t,s)=\sum_{j=1}^d P_j(s)e^{-\frac i\ve \int_s^t \alpha_j(u)du}$. The unitary adiabatic operator given by $V_\ve(t,s)=W(t,s)\Psi_\ve(t,s)$ approximates of the evolution operator $U_\ve(t,s)$ in the sense  
$$
\|U_\ve(t,s)-V_\ve(t,s)\|\leq \ve[C_1'+C_2'(t-s)].
$$
This is detailed in the Proof of Proposition \ref{Qlem}, Section \ref{secproofprop1.2}
\end{rem}

By the Duhamel principle, the solution of \eqref{20.2.1} is 
\begin{equation}
z(t) = U_{\vl}(t,0)z(0) +\frac{1}{i\ve}\int_0^t U_\vl(t,s)R_2(s, \vl) ds.
\end{equation}
Then we obtain from \eqref{42}, $\|z(0)\|\le 1$ and $\sup_{0\leq s\leq t\leq 1}\|U_\vl(t,s)\|\le C_0$ (see \eqref{estVU} or \eqref{49.1}), so that, uniformly in $0\le t\le 1$, 
\begin{equation}
\big\| z(t) -V_\vl(t,0)z(0)\big\| \le C\big( \ve +\lambda^2 + \frac1\ve\sup_{0\le s \le 1} \|R_2(s,\vl)\| \big),
\label{1.46}
\end{equation}
with the bound \eqref{1.36.2} on $R_2$. Recall the product structure \eqref{41} of $V_\vl(t,s)$, in which the phase term (complex phases) evolves quickly for small $\ve$ and where $W_\vl(t)$ satisfies \eqref{38}. As the projections $P_j(t,\vl)$ are not generally orthogonal, the operator $K_\vl(t)$ \eqref{1.38.1} is not  anti-selfadjoint, and thus $W_\vl(t)$ is not unitary. The size of $V_\vl(t)$ is thus not dictated by the phase term $\Psi_\vl(t,s)$ alone. We now compare $W_\vl(t)$ with the {\em unitary} operator $W(t)$ which is defined as the solution of
\begin{equation}
\label{1.47}
\partial_t W(t,s) = K(t) W(t,s),
 \quad W(s,s)=\bbbone\quad \mbox{with}\quad  K(t) = \sum_{j=1}^d [\partial_t P_j(t)] P_j(t).
\end{equation}

For later purposes, we recall here that since the spectrum of $A(t)$ is simple, we can make $W(t,s)$ explicit, following \cite{B}:

\begin{lem}[Berry phase] \label{berryphase} 
Given the smooth eigenbasis $\{\phi_k(t)\}_{1\leq k\leq d}$ of $A(t)$ and $1\leq j\leq d$,  $0\leq s,t$ we have 
\begin{equation}
\varphi_j(t,s):= W(t,s)\phi_j(s) = e^{i\xi_j(t,s)}\phi_j(t), \quad \mbox{where}\quad 
\xi_j(t,s) = i \int_s^t\bra\phi_j(u)| \partial_t\phi_j(u)\rangle du.
\label{xiint}
\end{equation} 
$\xi_j(t,s)$ is real, it is called the Berry phase. 
\end{lem}  



\bigskip

The operator $W(t,s)$ does not depend on $\ve$ nor on $\lambda$ and is unitary, $W(t,s)^*=W(t,s)^{-1}=W(s,t)$. We have
\begin{equation}
\partial_s  \Big( W_\vl(t,s)W(t,s)^{-1}\Big) = W_\vl(t,s)  \big[ K(s) - K_\vl(s) \big]W(t,s)^{-1},
\end{equation}
which gives upon integration,
\begin{equation}
\label{1.49}
W(t,s) -W_\vl(t,s) = \int_s^t W_\vl(t,u)\big[ K(u) - K_\vl(u) \big]  W(u,s) du.
\end{equation}
Next, we write
\begin{align}
\label{1.50}
K(t)-K_\vl(t) &= \sum_{j=1}^d \Big([\partial_t P_j(t)]P_j(t)-[\partial_t P_j(t,\vl)]P_j(t,\vl)\Big)\\
&= \sum_{j=1}^d [\partial_t P_j(t)] (P_j(t)-P_j(t,\vl))-(\partial_t P_j(t,\vl)-\partial_t P_j(t))P_j(t,\vl).
\nonumber
\end{align}
Using the estimates $\| P_j(t,\vl)-P_j(t)\|\le 	\frac{4\lambda^2}{\Delta_0}\|v\|^2_\infty \|\gamma\|_{L^1}$ (see \eqref{1.30}) and \eqref{1.36-1}, we obtain from \eqref{1.50} the bound 
\begin{equation}
\big\| K(u)-K_\vl(u)\big\| \le C\lambda^2\big(1+\frac1\ve|\gamma(u/\ve)|\big).
\label{1.52}
\end{equation}
Now, since $\|W(t,s)\|\le 1$
we obtain from \eqref{1.49}
\begin{equation}
\big\| W_\vl(t,s) - W(t,s) \big\| \le C\lambda^2 \int_s^t \|W_\vl(t,u)\|\big(1+\frac1\ve|\gamma(u/\ve)|\big) du.
\label{1.53.1}
\end{equation}
The usual Dyson series expansion based on \eqref{38} gives the bound
\begin{equation}
\|W_\vl(t,s)\|\le e^{\int_s^t \|K_\vl(u)\|du}\le e^{C(1+\lambda^2)(t-s)} e^{C\lambda^2}.
\label{1.58-1}
\end{equation}
To show the second inequality in \eqref{1.58-1}, we note that (see \eqref{1.34} and \eqref{1.38.1})
\begin{equation*}
\|K_\vl(u)\|\le 2d\max_j\|\partial_tP_j(t,\vl)\|\le C\big( 1 +\lambda^2+\frac{\lambda^2}{\ve}|\gamma(u/\ve)|\big),
\end{equation*}
so that $\int_s^t \|K_\vl(u)\|du\le C[(t-s)(1+\lambda^2)+\lambda^2\|\gamma\|_{L^1}]$. We combine \eqref{1.53.1} with \eqref{1.58-1} into 
\begin{equation}
\|W_\vl(t,s)-W(t,s)\|\le  C\lambda^2 e^{C(\lambda^2+t-s)}, \ \ \ \mbox{for any $s\le t$.}
\label{1.59-1}
\end{equation}
We now combine the bound \eqref{1.59-1} with \eqref{1.46}, 
\begin{multline}
\sup_{0\le t\le 1}\big\| z(t) -W(t)\Psi_\vl(t)z(0)\big\|\\
\le C\big( \ve + \lambda^2 + \frac{1}{\ve}\sup_{0\le s \le 1} \|R_2(s,\vl)\| \big) +\lambda^2 C\sup_{0\le s\le1}\|\Psi_\vl(s)\|,
\label{161}
\end{multline}
where we set for short $W(t)= W(t,0)$, $\Psi_\vl(t)=\Psi_\vl(t,0)$. Moreover, we show below, see \eqref{1.55.2} and \eqref{2.67}, that
\begin{equation}	
\|\Psi_\vl(s)\| \le 2d \max_j e^{\lambda^2 \|v\|^2_\infty [ \|\gamma\|_{L^1} +\frac{C_\gamma}{(m-1)(m-2){2^{m-2}}}]}\le C.
\end{equation}
Thus, from \eqref{1.36.2}, \eqref{1.29}, \eqref{R0} we get 
\begin{equation}
\frac{1}{\ve}\sup_{0\le s \le 1} \|R_2(s,\vl)\|
\leq C\big(\lambda^2+\frac{\lambda^4}{\ve}\big).
\label{163}
\end{equation}
Replacing in $\Psi_\vl(t)$, \eqref{41},  the projection $P_j(0,\vl)$ by $P_j(0)$ we incur an error of order $\lambda^2$ (see\eqref{1.30}) and thus finally have the following result from \eqref{161}-\eqref{163}:
\begin{equation}
\sup_{0\le t\le 1} \big\| z(t) -W(t) \sum_{j=1}^d P_j(0) e^{-\frac i\ve \int_0^t [\alpha_j(u)+\lambda^2\alpha'_j(u,\ve)]du} z(0) \big\|  \le C \Big( \ve + \lambda^2+ \frac{\lambda^4}{\ve}\Big),
\label{164.1}
\end{equation}
where $W(t)=W(t,0)$ is the solution of \eqref{1.47}.  The leading term in \eqref{164.1} can be simplified using Lemma \ref{berryphase}, 
\begin{align}
W(t)  P_j(0) z(0)=z_j(t)W(t)\phi_j(0)=
z_j(0) e^{i\xi_j(t)}\phi_j(t),
\end{align}
where we write for short $\xi_j(t)=\xi_j(t,0)$ for the Berry phase. Hence \eqref{164.1}  becomes
\begin{equation}
\sup_{0\le t\le 1} \big\| z(t) - \sum_{j=1}^d  e^{-\frac i\ve \int_0^t [\alpha_j(u)+\lambda^2\alpha'_j(u,\ve)]du} e^{i\xi_j(t)} z_j(0)\phi_j(t) \big\|  \le C \Big( \ve + \lambda^2+ \frac{\lambda^4}{\ve}\Big).
\label{164}
\end{equation}

We finally simplify the expression $e^{-i\frac{\lambda^2}{\ve} \int_0^t\alpha_j'(u,\ve)du }$  in \eqref{164}. We have from \eqref{1.43}
\begin{equation}
\int_0^t \alpha'_j(u,\ve) du = -i\int_0^t|v_j(u)|^2 \Big[\int_0^{u/\ve}e^{ix\alpha_j(u)}\gamma(x) dx\Big] du
\label{165}
\end{equation}
We first simplify the integral over $x$ by extending its upper bound to $\infty$. From \eqref{1.59.1},
\begin{align}\label{gamscat}
\left|\int_0^{u/\ve}e^{ix\alpha_j(u)}\gamma(x)dx-\int_0^{\infty}e^{ix\alpha_j(u)}\gamma(x)dx\right|\leq C_\gamma \int_{u/\ve}^\infty \frac{dx}{(1+x)^m}\leq \frac{C_\gamma}{(m-1)(1+u/\ve)^{m-1}},
\end{align}
and in view of \eqref{165},  $\int_0^t\frac{|v_j(u)|^2}{(1+u/\ve)^{m-1}}du \leq \ve \sup_{s\geq 0}|v_j(s)|^2  \int_0^\infty \frac{dy}{(1+y)^{m-1}}=C\ve$. Therefore, 
\begin{equation}
\Big|\int_0^t \alpha'_j(u,\ve) du  +i\int_0^t|v_j(u)|^2 \Big[\int_0^\infty e^{ix\alpha_j(u)}\gamma(x) dx\Big] du\Big| \le C\ve.
\label{166}
\end{equation}
Since $|e^z-e^{z'}| = |\int_z^{z'}e^\zeta d\zeta|\le C|z-z'|$ (where $C$ is the sup of $|e^\zeta|$ as $\zeta$ varies on the straight line joining $z$ and $z'$), we conclude that 
\begin{equation}
\Big| e^{-\frac{i\lambda^2}{\ve}\int_0^t\alpha'_j(u,\ve)du} - e^{-\frac{\lambda^2}{\ve} \int_0^t|v_j(u)|^2 \big[\int_0^\infty e^{ix\alpha_j(u)}\gamma(x)dx\big]du}
\Big|\le C\lambda^2,
\label{1.68}
\end{equation}
uniformly in $0\le t\le 1$. Next, as $\gamma(-x)=\overline\gamma(x)$, 
\begin{eqnarray*}
{\rm Re} \int_0^\infty e^{ix\alpha}\gamma(x)dx &=& \frac12\int_{\mathbb R} e^{ix\alpha}\gamma(x)dx =  \sqrt{\pi/2}\  \widehat\gamma(\alpha)\ge 0,\\
{\rm Im} \int_0^\infty e^{ix\alpha}\gamma(x)dx &=& \sqrt{2\pi}\  {\rm Im} (\widehat{\chi_+\gamma})(\alpha),
\end{eqnarray*}
where $\chi_+(x)=1$ for $x\ge 0$ and $\chi_+(x)=0$ else.

As $\|W(t)\|=1$, $\|P_j(0)\|=1$,  $|z(0)|\le 1$ and $\alpha_j(u)\in\mathbb R$, we can use \eqref{1.68} in \eqref{164} and we obtain Theorem \ref{leadingorder}.

\subsection{State of the emitted excitation and proof of Theorem \ref{thm1.9}}
\label{sect:emitted}

The average of a field observable $B$ (acting on $L^2(\rx^3,d^3k)$) in the state $\psi(t)$, \eqref{sol}, is
\begin{eqnarray}
\langle B\rangle_t\equiv \langle\psi(t), (\bbbone_\a\otimes B)\psi(t)\rangle &=& \|z(t)\|^2 \langle\Omega_\f, B\Omega_\f\rangle + \langle\Omega_\f, a(f_t)Ba^*(f_t)\Omega_f\rangle\nonumber\\
&=&\langle\Omega_\f, [a(f_t),B] a^*(f_t)\Omega_\f\rangle +\langle\Omega_\f,B\Omega_\f\rangle,
\end{eqnarray}
where we used that $a(f_t)a^*(f_t)\Omega_\f = \|f_t\|^2\Omega_\f = (1-\|z(t)\|^2)\Omega_\f$ (see \eqref{13}). Let us examine the probability density of the field excitation to have a specific fixed momentum $k_0\in\rx^3$. For this we take $B=a^*_{k_0}a_{k_0}$. Then $[a(f_t),a^*_{k_0}a_{k_0}]=\overline{f_t(k_0)}a_{k_0}$ and 
\begin{equation}
\langle a^*_{k_0}a_{k_0}\rangle_t = |f_t(k_0)|^2.
\end{equation}
In other words, $|f_t(k_0)|^2$ is the probability density of finding the field excitation in the  momentum $k_0$ at time $t$.

\subsubsection{ Proof of Theorem \ref{thm1.9}} 

The equation \eqref{14.1} with the initial condition $f_0=0$ yields
\begin{equation}
f_t(k)=-i\frac{\lambda}{\ve} g(k) \int_0^t \bra w(s)| z(s) \ket e^{-i\frac{(t-s)}{\ve}\omega(k)} ds.
\label{183}
\end{equation}
Then 
\begin{equation}
\langle B\rangle_t :=\int_{\rx^3} B(k)|f_t(k)|^2 d^3k = \frac{\lambda^2}{\ve^2} \int_0^t ds \int_0^t ds'\  \overline{\bra w(s)| z(s) \ket} \bra w(s')| z(s') \ket  \gamma_B\big(\frac{s-s'}{\ve}\big)
\label{185-1}
\end{equation}
where $\gamma_B$ is defined in Theorem \ref{thm1.9}. 
From Theorem \ref{leadingorder}, setting $\varphi_j(s)=e^{i\xi_j(s)}\phi_j(s)$,
\begin{multline}
\overline{\langle w(s), z(s)\rangle} \langle w(s'), z(s')\rangle = \big[\ \overline{\langle w(s), \varphi_j(s)\rangle} \  e^{\frac{i}{\ve}\int_0^s [\alpha_j+\lambda^2\widetilde\alpha_j]} e^{-\frac{\lambda^2}{\ve}\int_0^s \beta_j} +O\big(\ve+\lambda^2+\frac{\lambda^4}{\ve}\big)    \big]\\
\times \big[ \langle w(s'), \varphi_j(s')\rangle \  e^{-\frac{i}{\ve}\int_0^{s'} [\alpha_j+\lambda^2\widetilde\alpha_j]} e^{-\frac{\lambda^2}{\ve}\int_0^{s'} \beta_j} +O\big(\ve+\lambda^2+\frac{\lambda^4}{\ve}\big) \big]\\
 = \overline{\langle w(s), \varphi_j(s)\rangle} \langle w(s'), \varphi_j(s')\rangle\  e^{\frac{i}{\ve}\int_{s'}^s [\alpha_j+\lambda^2\widetilde\alpha_j]} e^{-\frac{\lambda^2}{\ve}[\int_0^{s'}+\int_0^s]\beta_j}
 + O\big(\ve+\lambda^2+\frac{\lambda^4}{\ve}\big).
 \label{186-1}
\end{multline}
The remainder term of \eqref{186-1}, inserted into \eqref{185-1}, gives 
\begin{equation}
\frac{\lambda^2}{\ve^2}O\big(\ve+\lambda^2+\frac{\lambda^4}{\ve}\big) \int_0^t ds \int_0^t ds' \gamma_B\big(\frac{s-s'}{\ve}\big) = O\big(\lambda^2 +\frac{\lambda^4}{\ve}+\frac{\lambda^6}{\ve^2} \big),
\end{equation}
because $0\le t\le 1$ and $|\int_0^t dx \gamma_B(x/\ve)|\le \ve \|\gamma_B\|_{L^1}$. It follows that 
\begin{eqnarray}
\langle B\rangle_t &=& \frac{\lambda^2}{\ve^2}\int_0^t ds \int_0^t ds' \overline{h(s)} h(s') \  e^{\frac{i}{\ve}\int_{s'}^s [\alpha_j+\lambda^2\widetilde\alpha_j]} e^{-\frac{\lambda^2}{\ve}[\int_0^{s'}+\int_0^s]\beta_j} \gamma_B\big(\frac{s-s'}{\ve}\big)\nonumber\\
&& +O\big(\lambda^2 +\frac{\lambda^4}{\ve}+\frac{\lambda^6}{\ve^2} \big),  \label{188}\\
h(s) &:=&\langle w(s), \varphi_j(s)\rangle. \nonumber
\end{eqnarray}
Morally, the oscillatory term in the integral has a phase $\sim \frac{s-s'}{\ve}$ while the decaying one scales as $\sim \frac{\lambda^2}{\ve}(s+s')$. This is why we are going to switch to the coordinates
$$
x=s-s'\in[-t,t],\quad y=s+s'\in[0,2t],
$$
and scale those separately. The square $[0,t]\times[0,t]$ to be integrated over in the variables $(s,s')$ becomes, in the $(x,y)$ plane, the square with one diagonal given by $x=0$ and $0\le y\le 2t$. Given $x\in[-t,t]$, the variable $y$ varies between $|x|\le y\le 2t- |x|$. Moreover, $s=\frac{y+x}{2}$, $s'=\frac{y-x}{2}$ and the Jacobian of the transformation is $\frac12$. Upon making the changes of variables $x/\ve\mapsto x$ and then $\frac{\lambda^2}{2\ve}y\mapsto y$, the main term on the right side of \eqref{188} is 
\begin{multline}
\frac{\lambda^2}{2\ve^2}\int_{-t}^t dx \int_{|x|}^{2t-|x|} dy \ \overline{h\Big(\frac{y+x}{2}\Big)} h\Big(\frac{y-x}{2}\Big) \gamma_B\Big(\frac{x}{\ve}\Big)
e^{\frac{i}{\ve}\int_{(y-x)/2}^{(y+x)/2} [\alpha_j+\lambda^2\widetilde\alpha_j]} e^{-\frac{\lambda^2}{\ve}[\int_0^{(y-x)/2}+\int_0^{(y+x)/2}]\beta_j}\\
= \frac{\lambda^2}{2\ve}\int_{-t/\ve}^{t/\ve} dx \gamma_B(x)\int_{\ve|x|}^{2t-\ve|x|} dy \ \overline{h\Big(\frac{y+\ve x}{2}\Big)} h\Big(\frac{y-\ve x}{2}\Big) 
e^{\frac{i}{\ve}\int_{(y-\ve x)/2}^{(y+\ve x)/2} [\alpha_j+\lambda^2\widetilde\alpha_j]} e^{-\frac{\lambda^2}{\ve}[\int_0^{(y-\ve x)/2}+\int_0^{(y+\ve x)/2}]\beta_j}\\
= \int_{-t/\ve}^{t/\ve} dx \gamma_B(x)\int_{\lambda^2|x|/2}^{\lambda^2t/\ve-\lambda^2|x|/2} dy \ \overline{h\Big(\frac{\ve}{\lambda^2}y+\frac{\ve}{2} x\Big)} h\Big(\frac{\ve}{\lambda^2}y-
\frac{\ve}{2} x\Big) 
e^{\frac{i}{\ve}\int_{ y\ve /\lambda^2-\ve x/2}^{ y\ve /\lambda^2+\ve x/2} [\alpha_j+\lambda^2\widetilde\alpha_j]}\\
\times e^{-\frac{\lambda^2}{\ve}[\int_0^{
 y\ve /\lambda^2-\ve x/2}+\int_0^{ y\ve /\lambda^2+\ve x/2}]\beta_j}\\
 \equiv \int_\rx dx \int_0^\infty dy F(x,y,t,\ve,\lambda),
\label{189-1}
\end{multline}
where ($\chi$ is the characteristic function)
\begin{multline}
F(x,y,t,\ve,\lambda) = \chi\big(|x|\le t/\ve\big)\,  \chi\Big(\lambda^2|x|/2\le y\le\lambda^2t/\ve-\lambda^2|x|/2\Big)\, \gamma_B(x) \\
\times \overline{h\Big(\frac{\ve}{\lambda^2}y+\frac{\ve}{2} x\Big)} h\Big(\frac{\ve}{\lambda^2}y-
\frac{\ve}{2} x\Big) 
e^{\frac{i}{\ve}\int_{ y\ve /\lambda^2-\ve x/2}^{ y\ve /\lambda^2+\ve x/2} [\alpha_j+\lambda^2\widetilde\alpha_j]}  e^{-\frac{\lambda^2}{\ve}[\int_0^{
 y\ve /\lambda^2-\ve x/2}+\int_0^{ y\ve /\lambda^2+\ve x/2}]\beta_j}.
 \label{190}
\end{multline}
Due to \eqref{wellc}, 
\begin{equation}
\label{wellcoupled}
\inf_{t\ge 0}\beta_j(t) \equiv \beta_j^* >0.
\end{equation}
In view of estimating the exponentially decaying term on the right side of \eqref{190}, we note that  
$$
\frac{\lambda^2}{\ve}\big[\frac{y\ve}{\lambda^2} -\frac{\ve x}{2}+  \frac{y\ve}{\lambda^2} +\frac{\ve x}{2}\big]=2y
$$
and thus $e^{-\frac{\lambda^2}{\ve}[\int_0^{
 y\ve /\lambda^2-\ve x/2}+\int_0^{ y\ve /\lambda^2+\ve x/2}]\beta_j}\le e^{-2y\beta_j^*}$. As $h$ is bounded, we get
\begin{equation}
|F(x,y,t,\ve,\lambda)| \le  C|\gamma_B(x)|\, e^{-2y\beta_j^*}. 
 \label{191}
\end{equation}

{\em $\bullet$ Consider first the regime (A).\ } The right hand side of \eqref{191} is integrable in $(x,y)\in \rx\times \rx_+$ and so by the dominated convergence theorem (recall \eqref{189-1}),
\begin{equation}
\lim_{\ve,\, \lambda,\, \ve/\lambda^2\, \rightarrow\, 0} \int_\rx dx \int_0^\infty dy \ F(x,y,t,\ve,\lambda) = \int_\rx dx \int_0^\infty dy \lim_{\ve,\, \lambda,\, \ve/\lambda^2\, \rightarrow\, 0}  F(x,y,t,\ve,\lambda).
\label{194}
\end{equation}
We now calculate the pointwise limit of $F$. We have 
\begin{equation}
\label{l1}
\lim_{\ve,\, \lambda,\, \ve/\lambda^2\, \rightarrow\, 0} h\Big(\frac{\ve}{\lambda^2}y\pm\frac{\ve}{2} x\Big) = h(0) =\langle w(0), \phi_j(0)\rangle
\end{equation}
and, denoting by $\mathcal A(t)$ the anti-derivative of $\alpha_j(t)+\lambda^2 \widetilde \alpha_j(t)$,
\begin{equation}
\frac{i}{\ve}\int_{ y\ve /\lambda^2-\ve x/2}^{ y\ve /\lambda^2+\ve x/2} [\alpha_j+\lambda^2\widetilde\alpha_j] = \frac{i}{\ve}\big[\mathcal A\Big(\frac{y\ve}{\lambda^2} + \frac{\ve x}{2}\Big) - \mathcal A\Big(\frac{y\ve}{\lambda^2} - \frac{\ve x}{2}\Big)\big] = \frac{i}{\ve} \ve x \big[\alpha_j(\tau)+\lambda^2\widetilde \alpha_j(\tau)\big],
\label{mvt}
\end{equation}
where we have used the mean value theorem in the last step, and where $\tau$, which depends on $x,y,\ve,\lambda$, satisfies $|\tau-\frac{y\ve}{\lambda^2}|\le \frac{\ve |x|}{2}$. It follows that 
\begin{equation}
\label{l2}
\lim_{\ve,\, \lambda,\, \ve/\lambda^2\, \rightarrow\, 0}  e^{\frac{i}{\ve}\int_{ y\ve /\lambda^2-\ve x/2}^{ y\ve /\lambda^2+\ve x/2} [\alpha_j+\lambda^2\widetilde\alpha_j] } = e^{ix\alpha_j(0)}. 
\end{equation}
Next, 
\begin{equation}
\label{l3}
\lim_{\ve,\, \lambda,\, \ve/\lambda^2\, \rightarrow\, 0}  e^{-\frac{\lambda^2}{\ve} \int_0^{
 y\ve /\lambda^2\pm\ve x/2}\beta_j} =
e^{-y\beta_j(0)},
\end{equation}
which follows from 
\begin{eqnarray*}
\lim_{\lambda^2/\ve\rightarrow\infty} -\frac{\lambda^2}{\ve}\int_0^{y\ve/\lambda^2}\beta_j &=&-y \lim_{r\rightarrow\infty} r \int_0^{1/r}\beta_j = -y\beta_j(0)\nonumber\\
\lim_{\ve,\, \lambda,\, \ve/\lambda^2\, \rightarrow\, 0} -\frac{\lambda^2}{\ve} \int_{y\ve/\lambda^2}^{y\ve/\lambda^2\pm \ve x/2}\beta_j & =& \lim_{\ve,\, \lambda,\, \ve/\lambda^2\, \rightarrow\, 0} \mp\frac{\lambda^2 x}{2}\beta_j(\tau)
=0,
\end{eqnarray*}
where we used the mean value theorem, as above in \eqref{mvt}, with $|\tau -\frac{y\ve}{\lambda^2}|\le \frac{\ve |x|}{2}$. We combine the limits \eqref{l1}, \eqref{l2}, \eqref{l3} into 
$$
\lim_{\ve,\, \lambda,\, \ve/\lambda^2\, \rightarrow\, 0}  F(x,y,t,\ve,\lambda) = \gamma_B(x)|\langle w(0),\varphi_j(0)\rangle|^2 \ e^{ix\alpha_j(0)}\, e^{-2y\beta_j(0)}.
$$
Note finally that $O(\lambda^2+\frac{\lambda^4}{\ve}+\frac{\lambda^6}{\ve^2})=O(\frac{\lambda^6}{\ve^2})$ in regime A.
Combining this with \eqref{194} and \eqref{189-1}, \eqref{188} yields
\begin{equation}
 \lim_{\ve,\, \lambda,\, \ve/\lambda^2,\, \lambda^3/\ve \, \rightarrow\, 0} \langle B\rangle_t = \sqrt{\frac{\pi}{2}}\, |\langle w(0),\phi_j(0)\rangle|^2\,  \frac{\widehat\gamma_B\big(\alpha_j(0)\big)}{\beta_j(0)}.
 \label{198}
\end{equation}
For the last equality, we use $\xi_j(0)=0$ (Berry phase) and due to \eqref{7}, $\langle w(0),\phi_j(0)\rangle = w_j(0)=v_j(0)$ and the definition of the Fourier transform,
$$
\widehat\gamma_B(\alpha) = \frac{1}{\sqrt{2\pi}}\int_\rx e^{it\alpha}\gamma_B(t)dt.
$$

{\em $\bullet$ Next consider the regime (B).\ } We can still use the dominated convergence theorem to calculate
\begin{equation}
\lim_{\ve,\, \lambda\rightarrow\, 0,\, \lambda^2/\ve=r\, } \int_\rx dx \int_0^\infty dy \ F(x,y,t,\ve,\lambda) = \int_\rx dx \int_0^\infty dy \lim_{\ve,\, \lambda\rightarrow\, 0,\, \lambda^2/\ve=r\, } F(x,y,t,\ve,\lambda).
\label{194'}
\end{equation}
Now the pointwise limit of $F$ is
\begin{equation}
\label{l1'}
\lim_{\ve,\, \lambda\rightarrow\, 0,\, \lambda^2/\ve=r\, }F(x,y,t,\ve,\lambda) = \chi\big(0\le y\le rt\big)\gamma_B(x) |h(y/r)|^2 e^{ix\alpha_j(y/r)} e^{-2r\int_0^{y/r}\beta_j},
\end{equation}
the error term is $O(\ve)$,
from which \eqref{limit'} follows at once. This concludes the proof of Theorem \ref{thm1.9}.\qed

\begin{rem} 
One can further prove the following estimate in the regime 
$\lambda^2\leq \ve$:
\begin{align}
\bra B \ket_{t}&=\frac{\lambda^2}{\ve}\sqrt{2\pi}\int_0^t  |v_j(s) |^2\hat\beta(\alpha_j(s))ds + O\Big(\lambda^2  + \frac{\lambda^4}{\ve^2}\Big),
\end{align}
further assuming $|\gamma_B(t)|\leq C_B (1+|t|)^{-\mu}$, with $\mu > 2$. 
\end{rem}

\subsection{The time independent case, proof of Corollary \ref{cor1.13}}
\label{timind}

We consider here the special case where the total Hamiltonian \eqref{1} is time independent, $H(t)\equiv H$. In keeping with the observation following \eqref{autonom}, we will set $\ve=1$ and consider $t\geq 0$. The time independent quantities appearing in the Hamiltonian are then denoted without the argument $t$: $A$, $w$, and so on.

In the time independent case, Propositions \ref{prop1} and \eqref{Qlem} reduce to the  following result.

\begin{prop}
\label{propti} Suppose the Hamiltonian \eqref{1} is time independent and the initial condition $\psi\in\mathcal P_1$ is such that $f_{t=0}=0$, so the excitation is initially in the atom. Then
	\begin{equation}
	\label{20.1ti}
	i \partial_tz(t) = \big[A -i\lambda^2 |w\ket\langle w| \Gamma(t)\big] z(t) + \tilde R_0(t, \lambda),
	\end{equation}
	where
	\begin{equation}
	\label{22ti}
	\Gamma(t) = \int_0^{t}e^{i xA}\, \gamma(x)dx,
	\end{equation}
and the remainder satisfies
	$$
	\sup_{t\geq 0}\|\tilde R_0(t, \lambda)\| \le \lambda^4\ \|w\|^4 \|\gamma\|_{L^1}\, \|t\gamma(t)\|_{L^1}.
	$$
\end{prop}
\medskip

The generator \eqref{20.2.2} of the linear approximation of $z(t)$ is replaced in the time independent case here by
\begin{align}
G_{\lambda}(t)=A -i\lambda^2 |w\ket\langle w| \Gamma(t),
\end{align}
with associated propagator $U_\lambda$, defined as the solution of 
\begin{align}\label{jula}
\i \partial_t U_\lambda(t,s)=G_\lambda(t) U_\lambda(t,s), \quad U_\lambda(s,s)=\bbbone.
\end{align}
As above, we will also write $U_\lambda(t)$ for $U_\lambda(t,0)$. Define
\begin{align}
\Gamma^+
=\int_0^\infty e^{ixA}\gamma(x) dx
\end{align}
Assuming the decay condition of $\gamma$, \eqref{1.59.1}, we have 
\begin{align}
\label{estdiffg}
\|\Gamma(t)-\Gamma^+\|\leq \int_t^\infty |\gamma(x)| dx\le  \frac{C_\gamma}{(m-1)(1+t)^{m-1}}\in L_1(\rx).
\end{align}

We adopt the well-coupledness assumption \eqref{wellc} modified to the time-independent setting,
\begin{equation}
\beta_{\rm min}\equiv \min_{1\leq j\leq d}\beta_j=\min_{1\leq j\leq d}\sqrt{\pi/2}|w_j|^2\widehat \gamma(\alpha_j)>0.
\label{wellcti}
\end{equation}

\begin{prop}
\label{scatprop}  Assume \eqref{wellcti}. For any $0<\beta^*<\beta_{\rm min}$,
there exists $\lambda_0>0$ and $C_0>0$ such that if $\lambda\leq  \lambda_0$ then the solution of \eqref{jula} satisfies  for all $t\ge 0$,  
\begin{equation}
\|U_\lambda(t)-e^{-it(A -i\lambda^2 |w\ket\langle w| \Gamma^+)}\|  \leq \lambda^2  \|w\|^2 C_0^2 e^{C_0\lambda^2\|w\|^2\|\Gamma-\Gamma^+\|_{L^1}}\, e^{-\beta^* t \lambda^2}\, \int_0^t  \|\Gamma(s)-\Gamma^+\|\, ds.
\label{enfin}
\end{equation}
Moreover, for all $t\geq s\geq  0$,
\begin{equation}
\| e^{-it(A -i\lambda^2 |w\ket\langle w| \Gamma^+)}\|\leq C_0e^{-\beta^* t \lambda^2}, \quad \| U_\lambda(t,s)\|\leq C_0e^{-\beta^* (t-s) \lambda^2}e^{C_0\lambda^2\|w\|^2\|\Gamma-\Gamma^+\|_{L^1}}.
\label{xxx}
\end{equation}
\end{prop}
\bigskip

It follows from \eqref{20.1ti}, the definition of $U_\lambda(t,s)$ given in \eqref{jula} and from the Duhamel formula, that 
\begin{equation}
z(t) = U_{\lambda}(t)z(0) -i\int_0^t U_\lambda(t,s)\tilde R_0(s, \lambda) ds.
\label{1.123}
\end{equation}
We estimate the remainder term employing \eqref{xxx},
\begin{equation}
\label{1.124}
    \left\|\int_0^t U_\lambda(t,s)\tilde R_0(s, \lambda) ds\right\| \leq C_R \lambda^4 \int_0^te^{-\beta^* (t-s) \lambda^2}ds
    \leq \frac{C_R}{\beta^*} \lambda^2\big(1-e^{-\beta^*t\lambda^2}\big),
\end{equation}
with $C_R= \|w\|^4 \|\gamma\|_{L^1} \|t\gamma(t)\|_{L^1} C_0e^{C_0\lambda^2\|w\|^2\|\Gamma-\Gamma^+\|_{L^1}}$. Next we use the estimate \eqref{enfin} to obtain
\begin{equation}
\label{xm1}
\| U_\lambda(t)z(0) -e^{-it(A-i\lambda^2|w\rangle\langle w|\Gamma^+)}z(0)\|\le C_1 \lambda^2 e^{-\beta^*t\lambda^2}\min(t,1),
\end{equation}
where $C_1 = \|w\|^2 C^2_0e^{C_0\lambda^2\|w\|^2\|\Gamma-\Gamma^+\|_{L^1}}(\|\Gamma-\Gamma^+\|_{L^\infty} + \|\Gamma-\Gamma^+\|_{L^1})$ and we have taken into account that 
\begin{equation*}
    \int_0^t  \|\Gamma(s)-\Gamma^+\|\, ds\leq \left\{ \begin{matrix} t \|\Gamma-\Gamma^+\|_{L^\infty} \\
    \|\Gamma-\Gamma^+\|_{L^1}
    \end{matrix}  \right.
    \le \big(\|\Gamma-\Gamma^+\|_{L^\infty} + \|\Gamma-\Gamma^+\|_{L^1} \big)\min(t,1).
\end{equation*}
Combining \eqref{1.123}, \eqref{1.124} and \eqref{xm1} yields the following result. 

\begin{cor}
\label{cor1.12}
Under the assumptions of Proposition \ref{scatprop},  for $\lambda\leq  \lambda_0$, and any $t\geq 0$,
\begin{align}
\| z(t)- e^{-it(A -i\lambda^2 |w\ket\langle w| \Gamma^+)}z(0)\| \leq 
C_2\lambda^2\left(\min (t,1)e^{-\beta^*t\lambda^2}+(1-e^{-\beta^*t\lambda^2})\right),
\label{1.126}
\end{align}
for $C_2=\|w\|^2 C_0^2e^{C_0\lambda^2\|w\|^2\|\Gamma-\Gamma^+\|_{L^1}}\{\frac{\|w\|^2}{\beta^*}\|\gamma\|_{L_1}\|t\gamma\|_{L^1} +C_0(\|\Gamma-\Gamma^+\|_{L^\infty} + \|\Gamma-\Gamma^+\|_{L^1})\}.$
\end{cor}

{\rm Remarks.\ }
i) By the Lumer-Phillips criterion, $e^{-it(A -i\lambda^2 |w\ket\langle w| \Gamma^+)}$ is a contraction semigroup if and only if
$-iA -\lambda^2 |w\ket\langle w| \Gamma^+$ is dissipative. However, the rank two operator
$$
{\rm Re} \, (-iA -\lambda^2 |w\ket\langle w| \Gamma^+)=-\frac{\lambda^2}{2}\big(|w\ket\langle w| \Gamma^++(\Gamma^+)^*|w\ket\langle w| \big)
$$
has positive and negative eigenvalues unless $w$ is an eigenvector of $(\Gamma^+)^*$, {\it i.e.} an eigenvector of $A$, which is forbidden by the well-coupledness assumption \eqref{wellcti}. In other words, there exists $z(0)$ of norm one such that for $t>0$ small enough, $\|e^{-it(A -i\lambda^2 |w\ket\langle w| \Gamma^+)}z(0)\|>1$, before the exponential decay kicks in. 

ii) The estimate \eqref{1.126} holds in particular for $t=0$ or $\lambda=0$, in which cases the approximation is trivially exact and the error term vanishes. 

iii) The error term in \eqref{1.126} is $O(\lambda^2)$ uniformly in $t\ge 0$.

\bigskip

We may diagonalize  $A-i\lambda^2|w\rangle\langle w|\Gamma^+ = \sum_{j=1}^d \alpha_j(\lambda)P_j(\lambda)$ and carry out analytic perturbation theory in $\lambda$, to obtain
\begin{equation}
 e^{-it(A -i\lambda^2 |w\ket\langle w| \Gamma^+)} =\sum_{j=1}^d e^{-it\alpha_j(\lambda) }P_j(\lambda)
 = \sum_{j=1}^d e^{-it\alpha_j(\lambda) }P_j + O\big(\lambda^2 e^{-\lambda^2t\beta_{\min}}\big).
 \label{1.127}
\end{equation}
provided $\lambda$ is small enough (see also \eqref{wellcti}). Here, $P_j$ is the eigenprojection of $A$ (associated to the eigenvalue $\alpha_j$) and 
\begin{equation}
\label{aaa}
\alpha_j(\lambda)=\alpha_j+\lambda^2\alpha'_j+O(\lambda^4),
\end{equation}
where $\alpha'_j\equiv \widetilde\alpha_j -i \beta_j$ is given in \eqref{aaa'}(see the proof of Proposition \eqref{scatprop}).

Combining \eqref{1.127} and \eqref{1.126}, 
\begin{align}
\sup_{t\ge 0}\| z(t)- \sum_{j=1}^d e^{-i t \alpha_j(\lambda)} P_jz(0)\| \leq 
\label{1.129}
C\lambda^2.
\end{align}
One further simplify the exponents by retaining only the the $O(\lambda^2)$ term in $\alpha_k(\lambda)$, according to \eqref{aaa}. Using that for any $\zeta\in\cx$, $|e^\zeta-1| = |\int_0^\zeta e^zdz|\le |\zeta| e^{|\zeta|}$,
\begin{eqnarray}
\big| e^{-i t\alpha_j(\lambda)} - e^{-i t (\alpha_j+\lambda^2\alpha'_j)}\big| &=& e^{-\beta_j t\lambda^2} \big| e^{-i tO(\lambda^4)}-1\big|\nonumber\\
&\le& e^{-\beta_j t\lambda^2} |tO(\lambda^4)| e^{|tO(\lambda^4)|}\nonumber\\
&\le& c t\lambda^4 e^{-\beta_{\min}t\lambda^2} e^{ ct \lambda^4}\nonumber\\
&\le& c t\lambda^4 e^{-\frac12\beta_{\min}t\lambda^2} \nonumber\\
&\le& \frac{2c}{e\beta_{\min}}\lambda^2.
\label{1.130}
\end{eqnarray}
In the third step, we used that $|O(\lambda^4)|\le c\lambda^4$ for some $c\ge 0$ and in the fourth step we took $\lambda^2<\frac{\beta_{\min}}{2c}$. We combine \eqref{1.129} and \eqref{1.130} into Corollary \ref{cor1.13}. 

The advantage of Corollary \ref{cor1.13} over Corollary \ref{cor1.12} is that the generator of the approximate evolution is simpler as it only contains energy corrections of $O(\lambda^2)$ and it is provided by a contraction semigroup. On the flip side, the remainder term in Corollary \ref{cor1.12} is better for small times (it vanishes at $t=0$) while in  Corollary \eqref{cor1.13} the remainder is only guaranteed to be $O(\lambda^2)$. 
\bigskip

{\bf Proof of Proposition \ref{scatprop}.\ }  We consider \eqref{jula} in the interaction picture with
\begin{align}\label{ginf}
G_\lambda^+=A -i\lambda^2 |w\ket\langle w| \Gamma^+
\end{align}
so that for $0\leq s\leq t$
\begin{align}\label{intint}
U_\lambda(t,s)=e^{-i(t-s) G_\lambda^+}-\lambda^2 \int_s^t e^{-i(t-r) G_\lambda^+} |w\ket\bra w|(\Gamma(s)-\Gamma^+) U_\lambda(r,s)dr.
\end{align}
The adaptation of the perturbation Lemma \ref{lem1} to the $\ve$ and time independent operator $G_\lambda^+$ in \eqref{ginf} yields
for $\lambda$ small,
\begin{align}
    e^{-itG_\lambda^+}=\sum_{j=1}^d e^{-it\alpha_j(\lambda)}P_j(\lambda),
\end{align}
where $\alpha_j(\lambda)=  \alpha_j +\lambda^2 \alpha_j' +O(\lambda^4)$ with $\alpha_j'=-i|w_j|^2 \int_0^\infty e^{ix\alpha_j}\gamma(x)dx$ and $P_j(\lambda)=P_j+O(\lambda^2)$ has norm bounded by 2. Here $\alpha_j$ and $P_j$ are the eigenvalues and eigenprojectors of $A$. The decomposition of $\alpha'_j$ in real and imaginary parts is (see also see \eqref{177}),
$$
\alpha'_j = \sqrt{2\pi} |w_j|^2 {\rm Im}\widehat{(\chi_+\gamma)}(\alpha_j)-i \sqrt{\pi/2} |w_j|^2  \widehat\gamma(\alpha_j)\equiv \widetilde\alpha_j -i \beta_j.
$$
Thus, for any $0<\beta^*<\min_{1\leq j\leq d} \beta_j$, there exists of $\lambda_0$ and $C_0$
such that if $|\lambda|\leq \lambda_0$ then  $\| e^{-it G_\lambda^+}\|\leq C_0e^{-\beta^* t \lambda^2}$ for all $t\geq 0$. Hence we deduce from \eqref{intint} by iteration, or via Gronwall Lemma, that under the same conditions, and for all $t\geq s \geq 0$, 
$$\|U_\lambda(t,s)\|\leq C_0e^{-\beta^* (t-s)\lambda^2}e^{C_0\lambda^2\|w\|^2\|\Gamma-\Gamma^+\|_{L^1}}.$$ 
Consequently, plugging these estimates in \eqref{intint} for $s=0$ yields \eqref{enfin}. 
This completes the proof of the proposition.
 \qed

\section{Proofs of auxiliary results}

\subsection{Proof of Proposition \ref{prop1}.}

We eliminate the free dynamics by changing the variables, 
\begin{equation}
\label{18}
y(t) = U_\ad(t)^{-1} z(t),\qquad 
h_t =  e^{i\frac{\omega t}{\varepsilon}}f_t
\end{equation}
where $U_\ad(t)$ is given in \eqref{19}. The equations resulting from \eqref{14} and \eqref{11} for the new variables $y(t)$, $h_t$ are,
\begin{eqnarray}
i\varepsilon\partial_t y(t) &=&\lambda \langle g,\emin h_t\rangle\,  U_\ad(t)^{-1}w(t)\label{20} \\
i\varepsilon \partial_t h_t &=& \lambda \langle w(t), U_\ad(t)y(t)\rangle\, \eplu g.
\label{21}
\end{eqnarray}
True to the Wigner-Weisskopf procedure, we now integrate the equation for $h_t$ and insert the result in the equation for $y(t)$. The initial condition, describing the setup of spontaneous radiative decay of the atom, satisfies
\begin{equation}
h_{t=0}(k)=f_{t=0}(k)=0.
\end{equation}
Thus equation \eqref{21} gives
\begin{equation}
h_t = -\frac{i\lambda}{\varepsilon}\int_0^t \langle w(s), U_\ad(s)y(s)\rangle e^{i\frac{\omega s}{\varepsilon}} ds\, g,
\end{equation}
which together with \eqref{20} leads to
\begin{equation}
\partial_t y(t) = -\frac{\lambda^2}{\varepsilon^2} \beta(t) \int_0^t \langle \beta(s), y(s)\rangle \, \gamma\big(\frac{t-s}{\varepsilon}\big)ds,
\label{24}
\end{equation}
where we have introduced
\begin{equation}
\label{25}
\beta(t) = U_\ad(t)^{-1} w(t)
\end{equation}
and the field correlation function $\gamma$ is given in \eqref{25.1}
Next we replace $y(s)$ in the integral on the right side of \eqref{24} by $y(t)$. More precisely, as
$$
\|\beta(u)\|_{\cx^2} = \|U_\ad(u)^{-1} w(u)\|_{\cx^d}=\|w(u)\|_{\cx^d} = \big[\sum_{j=1}^d |w_j(u)|^2\big]^{1/2} = \big[\sum_{j=1}^d|v_j(u)|^2\big]^{1/2} = \|v(u)\|_{\cx^d}
$$ 
and $\|y(s)\|\le 1$, the relation \eqref{24} implies
\begin{equation}
\|\partial_ty(t)\|\le\frac{\lambda^2}{\varepsilon^2} \|v\|^2_\infty\int_0^t |\gamma\big(\frac{t-s}{\varepsilon}\big)| ds\le \frac{\lambda^2}{\varepsilon} \|v\|_\infty^2 \|\gamma\|_{L^1}.
\end{equation}
where
\begin{equation}
\|v\|_\infty = \sup_{t\ge 0}\|v(t)\|_{\cx^d}=\sup_{t\ge 0} \big[\sum_{j=1}^d |v_j(t)|^2\big]^{1/2}.
\end{equation}
Therefore, for $t\ge s\ge 0$,
\begin{equation}
\|y(t)-y(s)\| \le \int_s^t \|\partial_u y(u)\| du \le \lambda^2 \|v\|_\infty^2 \|\gamma\|_{L^1} \frac{t-s}{\varepsilon}.
\end{equation}
Using this bound in \eqref{24} yields
\begin{equation}
\label{28}
\partial_t y(t) = -\frac{\lambda^2}{\varepsilon^2}\beta(t) \big\langle\int_0^t \beta(s) \overline\gamma\big(\frac{t-s}{\varepsilon}\big)ds, y(t)\big\rangle +T_1(t),
\end{equation}
where
\begin{equation}
\|T_1(t)\| \le   \frac{\lambda^4}{\varepsilon^2}\|v\|_\infty^4\|\gamma\|_{L^1} \int_0^t \frac{t-s}{\varepsilon} \ \big| \gamma\big(\frac{t-s}{\varepsilon}\big)\big| ds\le \frac{\lambda^4}{\varepsilon}\|v\|_\infty^4\|\gamma\|_{L^1}\, \|t\gamma(t)\|_{L^1}.
\end{equation}
The derivative of $\beta(s)=U_\ad(s)^{-1}w(s)$ is $\sim \varepsilon^{-1}$ due to the rapidly oscillating phases, $i\partial_s U_\ad(s)^{-1}$ is of order  $\varepsilon^{-1}$ (see \eqref{19}, \eqref{25}). So we replace only the part $w(s)$ by $w(t)$ within $\beta(s)$ in the integrand of \eqref{28}:
\begin{eqnarray}
\int_0^t \langle U_\ad(s)^{-1}w(s), y(t)\rangle\, \gamma\big(\frac{t-s}{\varepsilon}\big) ds &=& \int_0^t \langle U_\ad(s)^{-1}w(t), y(t)\rangle\, \gamma\big(\frac{t-s}{\varepsilon}\big) ds \nonumber\\
&& +T_2(t).
\label{30}
\end{eqnarray}
Using $\| w(t)-w(s)\| \le \int_s^t \|\partial_u w(u)\| du \le(t-s) \|\partial_tw\|_\infty$, with
\begin{equation}
\|\partial_tw\|_\infty \equiv \sup_{t\ge 0} \|\partial_tw(t)\|_{\cx^d}
\end{equation}
 we obtain
\begin{eqnarray}
|T_2(t)| &\le& \int_0^t \|w(t)-w(s)\|\, \|y(t)\|\ \big|\gamma\big(\frac{t-s}{\varepsilon}\big)\big|ds\nonumber\\
&\le& \|\partial_t w\|_\infty \int_0^t (t-s)\big|\gamma\big(\frac{t-s}{\varepsilon}\big)\big|ds \le\varepsilon^2 \|\partial_tw\|_\infty \ \|t\gamma(t)\|_{L^1}.
\label{31}
\end{eqnarray}
Combining \eqref{30}, \eqref{31} with \eqref{28} shows that  
\begin{equation}
\label{32}
\partial_t y(t) = -\frac{\lambda^2}{\varepsilon^2}\beta(t) \big\langle\int_0^t U_\ad(s)^{-1} w(t) \overline\gamma\big(\frac{t-s}{\varepsilon}\big)ds, y(t)\big\rangle +T_3(t),
\end{equation}
with
\begin{equation}
\|T_3(t)\|\le \frac{\lambda^4}{\varepsilon} \|v\|_\infty^4 \|\gamma\|_{L^1}\, \|t\gamma(t)\|_{L^1} +\lambda^2 \|\partial_tw\|_\infty \|w\|_\infty \|t\gamma(t)\|_{L^1}.
\end{equation}
In terms of the original variable $z(t)=U_\ad(t)y(t)$ (see \eqref{18}), \eqref{32} reads
\begin{equation}
\partial_tz(t) = -\frac{i}{\varepsilon}A(t)z(t) -\frac{\lambda^2}{\varepsilon} w(t) \langle Q_\varepsilon(t)w(t), z(t)\rangle +T_4(t),
\end{equation}
where $\|T_4(t)\|= \|T_3(t)\|$ and where $Q_\ve(t)$ is given in \eqref{22}.  This finishes the proof of Proposition \ref{prop1}. \qed

\subsection{Proof of Proposition \ref{Qlem}}\label{secproofprop1.2}

We introduce the atomic adiabatic evolution $V_\ve(t,s)$
\begin{equation} 
	\label{23.1}
	i \ve\partial_t V_\ve(t,s) = \big(A(t) +\i \ve K(t)\big) V_\ve(t,s),\qquad V_\ve(s,s)=\bbbone,
\end{equation}
where Kato's generator is
\begin{equation}
	K(t) = \sum_{j=1}^d [\partial_t P_j(t)] P_j(t)
\end{equation}
with 
$$
P_j(t) = |\phi_j(t)\rangle\langle\phi_j(t)|
$$ 
the spectral projections of $A(t)$. Since $A(t)$ is self-adjoint and $K(t)^*=-K(t)$ the solution $V_\ve(t,s)$ of \eqref{23.1} is unitary, $V_\ve(t,s)^*=V_\ve(t,s)^{-1}=V_\ve(s,t)$. Defining the $\ve$-independent Kato intertwining operator $W(t)$ as the solution of the $\ve$-independent equation
\begin{equation}
	\partial_t W(t,s) = K(t)W(t,s),\qquad W(s,s)=\bbbone
\end{equation}
one readily checks the relation (use the intertwining relation $W(t,s)P_j(s)=P_j(t)W(t,s)$) 
\begin{equation}
	\label{26.1}
	V_\ve(t,s) = W(t,s)\Phi_\ve(t,s),\qquad \Phi_\ve(t,s) = \sum_{j=1}^d P_j(s)e^{-\frac{i}{\ve}\int_s^t \alpha_j(u)du}.
\end{equation}
Both $W(t,s)$ and $\Phi_\ve(t,s)$ are unitary.  The point of this factorization of $V_\ve$ is to separate out the quickly varying ($\ve$ small) phase term $\Phi_\ve$. The adiabatic evolution $V_\ve(t,s)$ approximates $U_\ve(t,s)$,
\begin{equation}
\big\| V_\ve(t,s)-U_\ve(t,s) \big\| \le \ve \big[ C_1' +(t-s)C_2'\big],
	\label{27.1}
\end{equation}
where
\begin{eqnarray*}
C_1' &=& c\frac{d^2}{\Delta_0} \max_j \|\partial_t P_j\|_\infty\nonumber\\
C_2' &=& c\frac{d^2}{\Delta_0} \Big[ (1+d \|A\|_\infty) \max_j\|\partial_tP_j\|^2_\infty 
+ \max_j\|\partial_t^2P_j\|_\infty + \frac{\max_j|\partial_t\alpha_j|_\infty}{\Delta_0} \max_j\|\partial_t P_j\|_\infty
\Big],
\end{eqnarray*}
and where $c$ is a numerical constant. A proof \eqref{27.1} is rather standard. We present the details in Section \ref{sec:proofprop1.4}, \eqref{2.51}-\eqref{2.75}, in the more complicate setting where $V_\ve, U_\ve$ (\eqref{19}, \eqref{23.1}) are replaced by $V_\vl, U_\vl$ (\eqref{40}, \eqref{35}).

Approximating $U_\ve(t,s)$ in \eqref{22} by $V_\ve(t,s)$ we obtain from  \eqref{26.1}, \eqref{27.1}
\begin{multline}
\Big\| Q_\ve(t) - \frac1\ve \int_0^t \overline\gamma\big(\frac{t-s}{\ve}\big) V_\ve(t,s)ds \Big\| \\
\le  \ve C'_1  \frac1\ve \int_0^t \Big|\gamma\big(\frac{t-s}{\ve}\big)\Big| ds +\ve C'_2\frac{1}{\ve}\int_0^t \Big|\gamma\big(\frac{t-s}{\ve}\big)\Big| (t-s) ds \\
	\le \ve C'_1 \|\gamma\|_{L^1} +\ve^2 C'_2 \|t\gamma\|_{L^1}.
		\label{28.1}
\end{multline}
Next we use \eqref{26.1}, 
\begin{multline}
V_\ve(t,s) = \sum_j W(t,s)P_j(s) e^{-\frac{i}{\ve}\int_s^t\alpha_j(u)du} = \sum_j P_j(t)W(t,s) e^{-\frac{i}{\ve}\int_s^t\alpha_j(u)du} \\
=\sum_j P_j(t) e^{-\frac{i}{\ve}\int_s^t\alpha_j(u)du} +  \sum_j P_j(t)\big(\int_s^t \partial_u W(u,s)du \big) e^{-\frac{i}{\ve}\int_s^t\alpha_j(u)du}.
\end{multline}
Now $\|\int_s^t\partial_uW(u,s)du\|\le (t-s) \|K\|_\infty \le d(t-s)\max_j\|\partial_tP_j\|_\infty$. Then, 
\begin{multline}
\Big\| \frac1\ve \int_0^t \overline\gamma\big(\frac{t-s}{\ve}\big) V_\ve(t,s)ds - \sum_{j=1}^d P_j(t)\frac1\ve\int_0^t \overline\gamma\big(\frac{t-s}{\ve}\big)  e^{-\frac{i}{\ve}\int_s^t\alpha_j(u)du}ds  \Big\| \\
\le   d^2\max_j\|\partial_tP_j\|_\infty 
\frac1\ve \int_0^t | \overline\gamma\big(\frac{t-s}{\ve}\big)|(t-s)ds \le \ve d^2\max_j\|\partial_tP_j\|_\infty \|t\gamma\|_{L^1}.
\label{2.26}
\end{multline}
Combining \eqref{28.1} and  \eqref{2.26} yields
\begin{multline}
	\Big\| Q_\ve(t) - \sum_{j=1}^d P_j(t)\frac1\ve \int_0^t \overline\gamma\big(\frac{t-s}{\ve}\big) e^{-\frac{i}{\ve}\int_s^t\alpha_j(u)du} ds \Big\| \\
	\le \ve  \big(C_1' \|\gamma\|_{L^1} +d^2\max_j\|\partial_tP_j\|_\infty \|t\gamma\|_{L^1}\big)  +\ve^2 C_2' \|t\gamma\|_{L^1}.
		\label{31.2}
\end{multline}
Next, we have 
\begin{equation}
	\label{33}
	\frac1\ve \int_0^t   \overline\gamma\big(\frac{t-s}{\ve}\big) e^{-\frac{i}{\ve}\int_s^t\alpha_j(u)du}ds = \int_0^{t/\ve}   \overline\gamma(x) e^{-\frac{i}{\ve}\int_{t-\ve x}^t\alpha_j(u)du}dx
\end{equation}
and since $|e^{ia}-e^{ib}|= |\int_a^b e^{iy}dy|\le |b-a|$,
\begin{eqnarray}
	\big| e^{-\frac{i}{\ve}\int_{t-\ve x}^t\alpha_j(u) du} - e^{-ix\alpha_j(t)}\big| &\le& \frac{1}{\ve}\int_{t-\ve x}^t | \alpha_j(u) -\alpha_j(t) |du \nonumber\\
	&\le& \|\partial_t\alpha_j\|_\infty \frac1\ve \int_{t-\ve x}^t (t-u)du = \frac12 \ve x^2 \|\partial_t\alpha_j\|_\infty.
\end{eqnarray}
Combining this with \eqref{33} and \eqref{31.2} gives
\begin{equation}
	\Big|\frac1\ve \int_0^t   \overline\gamma\big(\frac{t-s}{\ve}\big) e^{-\frac{i}{\ve}\int_s^t\alpha_j(u)du}ds - \int_0^{t/\ve} \overline\gamma(x) e^{-i x\alpha_j(t)}  dx\Big| \le \frac12 \ve \|\partial_t\alpha_j\|_\infty\,  \|t^2\gamma\|_{L^1}
\end{equation}
and finally
\begin{multline}
	\Big\| Q_\ve(t) - \sum_{j=1}^d P_j(t) \int_0^{t/\ve} e^{-ix\alpha_j(t)} \gamma(x)  dx \Big\| \le \ve  \big(C'_1 \|\gamma\|_{L^1} +d^2\max_j\|\partial_tP_j\|_\infty \|t\gamma\|_{L^1}\big)\\
	+\ve^2 C_2' \|t\gamma\|_{L^1} + \frac{d}{2} \ve \|\partial_t\alpha_j\|_\infty\,  \|t^2\gamma\|_{L^1}.\quad
\end{multline}
This is the result \eqref{28.2}, showing Proposition \ref{Qlem}. \qed

\subsection{Proof of Lemma \ref{lem1}} (a)  For $z\in\cx$ in the resolvent set of $A(t)$, we have
\begin{equation}
	\big\| (A(t)-z)^{-1}\big\| \le \frac{1}{{\rm dist}({\rm spec}(A(t)),z)}.
\end{equation}
Take $z\in \cx$ which is separated from ${\rm spec}(A(t))$ by at least some distance $a>0$. Then for $\lambda^2\|v\|_\infty^2 \|\gamma\|_{L^1}/a<1$, the resolvent $(G_\vl(t)-z)^{-1}$ is a bounded operator, given by the convergent Neumann series 
\begin{equation}
	\label{neumann}
	(G_\vl(t))-z)^{-1}= (A(t)-z)^{-1}\sum_{n\ge 0}\Big[ i\lambda^2  |w(t)\rangle\langle w(t)|\, \Gamma_\ve(t)\, (A(t)-z)^{-1}\Big]^n.
\end{equation}
Hence the spectrum of $G_\vl(t)$ lies in a neighbourhood of the size $\lambda^2\|v\|^2_\infty\|\gamma\|_{L^1}$ of the spectrum of $A(t)$. This shows that the eigenvalues of $G_\vl(t)$ are simple, and $\big| \alpha_j(t,\vl)-\alpha_j(t)\big|\le  \lambda^2\|v\|^2_\infty \|\gamma\|_{L^1}$. We will prove the bound \eqref{1.42} below, after analyzing the spectral projections. 

(b) Let $\C_j(t)$ be a circle around $\alpha_j(t)$, with radius $\Delta_0/2$ (c.f. \eqref{gap}). Since due to \eqref{lambdasmall} we have $\lambda^2\|v\|^2_\infty \|\gamma\|_{L^1} < \Delta_0/4$ the contour $\C_j(t)$ lies in the resolvent set of $G_\vl(t)$ and the Riesz projection 
\begin{equation}
	\label{1.35}
	P_j(t,\vl)= \frac{-1}{2\pi i}\int_{\C_j(t)} (G_\vl(t)-z)^{-1} dz
\end{equation}
is well defined. Next, 
\begin{eqnarray}
	\big\| P_j(t,\vl)-P_j(t)\big\| &\le& \frac{\lambda^2}{2\pi}\Big\| \int_{\C_j(t)} (G_\vl(t)-z)^{-1} |w(t)\rangle\langle w(t)| \Gamma_\ve(t)(A(t)-z)^{-1}dz\Big\|\nonumber\\
	&\le& \tfrac12\lambda^2 \Delta_0 \|v\|_\infty^2 \|\gamma\|_{L^1} \max_{z\in\C_j(t)} \|(G_\vl(t)-z)^{-1}\| \ 2\Delta_0^{-1}.
	\label{1.36}
\end{eqnarray}
From \eqref{neumann}, we obtain the following bound  for all $z\in\C_j(t)$, 
\begin{equation}
	\label{1.36.1}
	\|(G_\vl(t)-z)^{-1}\|\le 2\Delta_0^{-1}\sum_{n\ge 0} [\lambda^2 2 \Delta_0^{-1}\|v\|^2_\infty\|\gamma\|_{L^1}]^n\le 4\Delta_0^{-1}.
\end{equation}
Combining this bound with \eqref{1.36} gives $\big\| P_j(t,\vl)-P_j(t)\big\| <1$. Thus the ranks of $P_j(t,\vl)$ and $P_j(t)$ are equal, namely one. It follows that $G_\ve(t)$ has a single, simple eigenvalue inside $\C_j(t)$, with associated Riesz projection \eqref{1.35}. This shows (b). 

We give a proof of (c) now. From \eqref{1.35} and \eqref{1.36.1} we have $\|P_j(t,\vl)\|\le\frac{\Delta_0}{2}4\Delta_0^{-1}=2$. Next, from \eqref{1.35}, 
\begin{eqnarray}
	\partial_t P_j(t,\vl) &=&\frac{-1}{2\pi i}\  \int_{\C_j(t)} \partial_t(G_\vl(t)-z)^{-1} dz\nonumber\\
	&=&\frac{1}{2\pi i}\ \int_{\C_j(t)} (G_\vl(t)-z)^{-1}\big( \partial_t G_\vl(t)\big)  (G_\vl(t)-z)^{-1}dz.
	\label{1.54}
\end{eqnarray}
It is not necessary to consider the $t$-derivative of the curve $\C_j(t)$ as this curve can be taken constant in $t$ for $t$ in a neighbourhood of the point where the derivative is taken. From \eqref{20.2.2},
\begin{equation}
	\partial_t G_\vl(t) = \partial_tA(t) -2i\lambda^2 {\rm Re}\big( |\partial_t w(t)\rangle\langle w(t)|\big)\Gamma_\ve(t) -i\lambda^2 |w(t)\rangle\langle w(t)|\partial_t\Gamma_\ve(t).
	\label{1.55}
\end{equation}
From \eqref{1.24},
\begin{equation}
	\partial_t \Gamma_\ve(t) = \frac 1\ve \, e^{i\frac t\ve A(t)}\, \gamma(t/\ve) + \int_0^{t/\ve} \big( \partial_t e^{ixA(t)}\big)\gamma(x)dx
\label{2.37}
\end{equation}
and using that
\begin{equation}
	\partial_t e^{ixA(t)} = \int_0^1 e^{irx A(t)}\big(ix\partial_tA(t)\big) e^{i(1-r)xA(t)} dr
\label{2.38}
\end{equation}
we obtain
\begin{equation}
	\|\partial_t \Gamma_\ve(t)\| \le \frac1\ve |\gamma(t/\ve)| + \|t\gamma(t)\|_{L^1}\, \|\partial_tA\|_\infty.
	\label{1.58}
\end{equation}
We use the bound \eqref{1.58} in \eqref{1.55},
\begin{equation}
	\|\partial_t G_\vl(t)\| \le \|\partial_tA\|_\infty + 2\lambda^2\|\partial_t w\|_\infty\|w\|_\infty \|\gamma\|_{L^1} + \lambda^2\|v\|_\infty^2\big[ \frac1\ve |\gamma(t/\ve)| + \|t\gamma(t)\|_{L^1}\, \|\partial_tA\|_\infty\big].
	\label{1.45.1}
\end{equation}
Now we use this to bound $\|\partial_t P_j(t,\vl)\|$ as per \eqref{1.54},
\begin{equation}
	\|\partial_t P_j(t,\vl)\| \le \frac{\Delta_0}{2} \|\partial_t G_\vl(t)\| \max_{z\in\C_j(t)}\|(G_\vl(t)-z)^{-1}\|^2 \le\frac{8}{\Delta_0}\|\partial_tG_\vl(t)\|,
\end{equation}
where we have also employed \eqref{1.36.1}. Combining the last bound with \eqref{1.45.1} yields the desired \eqref{1.34}. 

Our next task is to show \eqref{1.36-1}. From \eqref{1.54},
\begin{multline}
\|\partial P_j(t,\vl) - \partial_t P_j(t) \|=\frac{1}{2\pi}\|\int_{\mathcal C_j(t)}\Big[(G_\vl(t)-z)^{-1}\big( \partial_t G_\vl(t)\big)  (G_\vl(t)-z)^{-1}\\
-(A(t)-z)^{-1}\big(\partial_tA(t)\big) (A(t)-z)^{-1}\Big]dz\|\\
\le \frac{c}{\Delta_0}\Big[\lambda^2 \|w\|_\infty^2\|\gamma\|_{L^1} \|\partial_t A\|_\infty  + \|\partial_tG_\vl(t)-\partial_tA(t)\|\Big],
\label{2.44-1}
\end{multline}
where $c$ is a numerical constant. To get this estimate we used the identity (dropping the variables and subscripts)
\begin{align}
&(G-z)^{-1}\big( \partial_t G\big)  (G-z)^{-1}
-(A-z)^{-1}\big(\partial_tA\big) (A-z)^{-1}\nonumber\\
&=
(G-z)^{-1}\big(\partial_tG -\partial_tA\big) (G-z)^{-1} + \big((G-z)^{-1}-(A-z)^{-1}\big)\big(\partial_tA\big)(G-z)^{-1}\nonumber\\
&+(A-z)^{-1}\big(\partial_tA\big)\big((G-z)^{-1}-(A-z)^{-1}\big),
\end{align}
 $\|(A(t)-z)^{-1}\|=2/\Delta_0$, $\|(G_\vl(t)-z)^{-1}\|\le 4/\Delta_0$ (see \eqref{1.36.1}) and (see \eqref{neumann})  $\max_{z\in\mathcal C_j(t)} \|(G_\vl(t)-z)^{-1}-(A(t)-z)^{-1}\|\le \frac{4\lambda^2}{\Delta_0}\|w\|_\infty^2\|\gamma\|_{L^1}$. Now from \eqref{1.55} and \eqref{1.58}, 
$$
\|\partial_tG_\vl(t) -\partial_t A(t)\|\le 2\lambda^2\|w\|_\infty\|\partial_t w\|_\infty\|\gamma\|_{L^1} +\lambda^2\|w\|^2_\infty \big(\frac1\ve|\gamma(t/\ve)| +\|t\gamma\|_{L^1}\|\partial_tA\|_\infty\big).
$$
Combining this with \eqref{2.44-1} yields \eqref{1.36-1}, with a constant $C$ we can make explicit if need be.

\medskip
We now bound the second derivative, $\partial_t^2P_j$. From \eqref{1.54}, and simply writing $G$ for $G_\vl(t)$, 
\begin{multline}
\partial_t^2 P_j(t,\vl) 
	=\frac{-1}{2\pi i}\ \int_{\C_j(t)} (G-z)^{-1}\big[ 2(\partial_t G)  (G-z)^{-1} (\partial_t G) - \partial_t^2G\big] (G-z)^{-1}dz.
	\label{2.42}
\end{multline}
Now by \eqref{1.36.1} and \eqref{1.45.1}
\begin{multline}
\|(G-z)^{-1} 2(\partial_t G)  (G-z)^{-1} (\partial_t G) (G-z)^{-1}\| \le \frac{128}{\Delta_0^3} \|\partial_tG\|^2\\
\le C\Big( 1+\lambda^2+\frac{\lambda^2}{\ve}|\gamma(t/\ve)|\big(1+\frac{\lambda^2}{\ve}|\gamma(t/\ve)|\big)\Big)
\label{2.43}
\end{multline}
for a (traceable) constant $C$ independent of $\vl$ and $t\geq 0$. Next,
$\|(G-z)^{-1} (\partial_t^2G)   (G-z)^{-1} \| \le   C\|\partial_t^2G\|$. We apply $\partial_t$ to \eqref{1.55} and obtain
\begin{eqnarray}
\| \partial_t^2G(t) \| &\leq & \|\partial_t^2A\|_\infty + 4\lambda^2 \big( \|w\|_\infty\|\partial_t^2w\|_\infty +\|\partial_t w\|_\infty^2\big) \|\gamma\|_{L^1}  \nonumber\\
&& + 2\lambda^2\|w\|_\infty\|\partial_t w\|_\infty \|\partial_t\Gamma(t)\| +\lambda^2 \|w\|_\infty^2 \|\partial_t^2\Gamma(t)\|.
\label{2.44}
\end{eqnarray}
The first derivative of $\Gamma(t)$ is estimated in \eqref{1.58}. Next, from \eqref{2.37}, \eqref{2.38},
\begin{eqnarray}
\partial^2_t\Gamma_\ve(t) &=&\frac1\ve \gamma(t/\ve) \int_0^1 e^{i\frac{r}\ve tA(t)} \frac{i}{\ve} \partial_t(tA(t)) e^{i\frac{1-r}{\ve} tA(t)}dr +  \frac{1}{\ve^2} e^{i\frac{t}{\ve} A(t)}(\partial_t\gamma)(t/\ve) \nonumber\\
&&+ \frac1\ve \gamma(t/\ve) \big(\partial_t e^{ixA(t)}\big)|_{x=t/\ve} + \int_0^{t/\ve}\big(\partial_t^2 e^{ixA(t)}\big)\gamma(x) dx.
\label{2.45}
\end{eqnarray}
To estimate $\partial^2_t e^{ixA(t)}$, we take  $\partial_t$ on both sides of \eqref{2.38} and use the unitarity of $e^{iyA(t)}$, any $y\in\rx$, to get
\begin{equation*}
\|\partial_t^2 e^{ixA(t)}\| \le  2 x^2\|\partial_tA\|_\infty^2 + x\|\partial_t^2A\|_\infty.
\end{equation*}
We then get the following upper bound on \eqref{2.45}, 
\begin{eqnarray}
\|\partial^2_t\Gamma_\ve(t)\| &\le &\frac{1}{\ve^2} |\gamma(t/\ve)|\big( \|A\|_\infty + t\|\partial_tA\|_\infty\big)   +  \frac{1}{\ve^2} |(\partial_t\gamma)(t/\ve)| \nonumber\\
&&+ \frac{t}{\ve^2} |\gamma(t/\ve)|\, \|\partial_tA\|_\infty  +   2 \|\partial_tA\|_\infty^2 \|x^2\gamma(x)\|_{L^1}+ \|\partial_t^2A\|_\infty \|x\gamma(x)\|_{L^1}.
\label{2.46}
\end{eqnarray}
Combining this with \eqref{2.44} yields
\begin{equation}
\|\partial^2_t G(t)\| \le C\big(1+\lambda^2 +\frac{\lambda^2}{\ve^2} (|\gamma(t/\ve)|(1+t) +|(\partial_t\gamma)(t/\ve)| \big) 
\label{2.47}
\end{equation}
for a constant $C$ independent of $\vl$ and $t\geq 0$. Finally, we combine the estimates \eqref{2.42}, \eqref{2.43} \eqref{2.47} to arrive at \eqref{1.35.3}.
\medskip

We now show \eqref{1.42}. We have $G_\vl(t)P_j(t,\vl)=\alpha_j(t,\vl)P_j(t,\vl)$. Since  $P_j(t,\vl)$ is a rank-one projection it has unit trace, and so 
\begin{equation}
	\alpha_j(t,\vl)= {\rm tr} \ G_\vl(t)P_j(t,\vl).
\end{equation}
We use $G_\vl(t)(G_\vl(t)-z)^{-1}= \bbbone +z(G_\vl(t)-z)^{-1}$
and get from \eqref{1.35},
\begin{equation}
	\alpha_j(t,\vl)=  {\rm tr}\,\frac{-1}{2\pi i}\oint_{\C_j} z\,  (G_\vl(t)-z)^{-1} dz.
	\label{1.35.1}
\end{equation}
Now we expand the resolvent, 
\begin{eqnarray}
	(G_\vl(t)-z)^{-1} &=& (A(t)-z)^{-1} +i\lambda^2(A(t)-z)^{-1}  |w(t)\rangle\langle w(t)|\Gamma_\ve(t)  (A(t)-z)^{-1} \nonumber\\
	&&+ (G_\vl(t)-z)^{-1} \big[ i\lambda^2 |w(t)\rangle\langle w(t)| \Gamma_\ve(t) (A(t)-z)^{-1} \big]^2
\end{eqnarray}
and inserting this into the integral on the right side of \eqref{1.35.1},
\begin{eqnarray}
	\lefteqn{\alpha_j(t,\vl) = {\rm tr}\, \frac{-1}{2\pi i}\oint_{\C_j(t)} z (A(t)-z)^{-1} dz}\nonumber\\
	&& +i\lambda^2 \frac{-1}{2\pi i}\oint_{\C_j(t)} z \, {\rm tr}\, (A(t)-z)^{-1} |w(t)\rangle\langle w(t)|\Gamma_\ve(t) (A(t)-z)^{-1} dz\nonumber\\
	&& +T_j(t,\vl).
	\label{1.37}
\end{eqnarray}
The remainder term is estimated as follows. Use that for any  $d\times d$ matrix $X$ one has $|{\rm tr} X|\le {\rm tr}|X|\le d\|X\|$ and that  $|\C_j(t)|=2\pi \Delta_0/2$, $\|(A(t)-z)^{-1}\| \le2/\Delta_0$, $\|\Gamma_\ve(t)\|\le \|\gamma\|_{L^1}$ and the bound $\| (|w(t)\rangle\langle w(t)|)\| \le \|v\|_\infty^2$, as well as \eqref{1.36.1}, and $|z|\le |\alpha_j(t)| + \Delta_0/2\le \|A\|_\infty+\Delta_0/2$. Then
\begin{equation}
	\label{1.38}
	|T_j(t,\vl)| \le d\frac{8}{\Delta_0^2} \lambda^4\|v\|^4_\infty \|\gamma\|^2_{L^1}(\|A\|_\infty + \Delta_0/2).
\end{equation}
The first term on the right side of \eqref{1.37} equals the trace of  $\alpha_j(t) P_j(t)$ (eigendata of $A(t)$), which is just $\alpha_j(t)$. Next, using that the $P_\ell(t)$ are a complete set of orthonormal projections, 
\begin{equation}
	{\rm tr}\, (A(t)-z)^{-1} |w(t)\rangle\langle w(t)|\Gamma_\ve(t) (A(t)-z)^{-1} =\sum_{\ell=1}^d \frac{ \|P_\ell(t)w(t)\|^2}{(\alpha_\ell(t)-z)^2} \int_0^{t/\ve}e^{ix\alpha_\ell(t)}\gamma(x)dx.
\end{equation}
In view of \eqref{1.37} we need to take the integral
\begin{equation}
	\label{1.41}
	\frac{-1}{2\pi i} \oint_{\C_j(t)}\frac{z}{(\alpha_\ell(t)-z)^2} dz  =-\delta_{j,\ell}
\end{equation}
(Kronecker symbol). Using that $\|P_\ell(t)w(t)\| =|v_\ell(t)|^2$ and combining \eqref{1.37}, \eqref{1.38} and \eqref{1.41} gives the bound \eqref{1.42}, \eqref{1.43}.

This completes the proof of Lemma \ref{lem1}.\qed

\subsection{Proof of Proposition \ref{prop4}}
\label{sec:proofprop1.4}

To verify \eqref{41} we take $i\ve\partial_t$ of the right side, using \eqref{38} and \eqref{41}:
\begin{eqnarray}
	\lefteqn{
		i\ve\partial_t \big( W_\vl(t,s)\Psi_\ve(t,s) \big)= i\ve K_\vl(t) W_\vl(t,s)\Psi_\vl(t,s)} \nonumber\\
	&& + W_\vl(t,s)\sum_{j=1}^d [\alpha_j(t)+\lambda^2 \alpha_j'(t,\ve)] P_j(s,\vl)e^{-\frac i\ve \int_s^t  [\alpha_j(u)+\lambda^2 \alpha_j'(u,\ve)] du}.
	\label{43}
\end{eqnarray}
Next, using the disjointness and completeness of the spectral projections, one readily sees that $\Psi_\vl(t,s)^{-1} = \sum_{j=1}^d P_j(s,\vl)e^{\frac i\ve\int_s^t [\alpha_j(u)+\lambda^2\alpha_j'(u,\ve)]du}$. Multiplying the last operator in \eqref{43} on the right by $\bbbone = \Psi_\vl(t,s)^{-1} \Psi_\vl(t,s)$ and using that 
\begin{eqnarray*}
\sum_{j=1}^d [\alpha_j(t)+\lambda^2\alpha'_j(t,\ve)] P_j(s,\vl)e^{-\frac i\ve \int_s^t [\alpha_j(u)+\lambda^2\alpha'_j(u,\ve)]du}\  \Psi_\vl(t,s)^{-1}\\
= \sum_{j=1}^d[\alpha_j(t)+\lambda^2\alpha'_j(t,\ve)]P_j(s,\vl),
\end{eqnarray*}
together with the intertwining property $W_\vl(t,s)P_j(s,\vl)= P_j(t,\vl)W_\vl(t,\vl)$ and the fact that $\sum_{j=1}^d [\alpha_j(t)+\lambda^2\alpha'_j(t,\ve)] P_j(t,\vl) = \G_\vl(t)$ ({\em c.f.} \eqref{1.35.2}), we see that the second term on the right side of \eqref{43} satisfies
\begin{eqnarray}
	\label{44}
	W_\vl(t,s)\sum_{j=1}^d [\alpha_j(t)+\lambda^2\alpha'_j(t,\ve)] P_j(s,\vl)e^{-\frac i\ve \int_s^t [\alpha_j(u)+\lambda^2\alpha'_j(u,\ve)]du} \nonumber\\
	= \G_\vl(t)W_\vl(t,s)\Psi_\vl(t,s).
\end{eqnarray}
Combining \eqref{44} with \eqref{43} shows that $W_\vl(t,s)\Psi_\vl(t,s)$ satisfies the same differential equation and the same initial condition as $V_\vl(t,s)$, \eqref{40}, so the two are equal. This shows \eqref{41}.

To prove \eqref{42} we proceed in a standard fashion following \cite{ASY}, see also \cite{J2, AFGG, J}. For the remainder of this proof, we omit from the notation the dependence of operators on $\vl$. That is, we simply write $K(t)$, $P_j(t)$, $\G(t)$, $V(t,s)$, $U(t,s)$ instead of  $K_\vl(t)$, $P_j(t,\vl)$, $\G_\vl(t)$, $V_\vl(t,s)$, $U_\vl(t,s)$ (the latter quantities having been defined in \eqref{1.38.1}, Lemma \ref{lambdasmall}(b), \eqref{1.35.2}, \eqref{40}, \eqref{35} respectively). 

We have 
\begin{equation}
	K(t) = \frac12\sum_{j=1}^d[ \partial_tP_j(t),P_j(t)] 
		\label{2.51}
\end{equation}
and for any operator $X$,
\begin{equation}
 [X, P_j(t)] = [\G(t),\R_j(t,X)],
\end{equation}
where 
\begin{equation}
\R_j(t,X) \equiv \sum_{\ell=1,\ldots,d,\, \ell\neq j} \frac{P_j(t)XP_\ell(t) + P_\ell(t) X P_j(t)}{\widetilde\alpha_\ell(t)-\widetilde\alpha_j(t)}.
\label{2.53}
\end{equation}
In \eqref{2.53}, $\widetilde\alpha_j(t) = \alpha_j(t)+\lambda^2\alpha_j'(t,\ve)$ are  the eigenvalues of $\G(t)$, see \eqref{1.35.2}. Integrating the relation 
\begin{equation*}
\partial_r\big[ V(t,r) U(r,s)\big] = -V(t,r)K(r) U(r,s)
\end{equation*}
over the domain $r$ ranging from $s$ to $t$, and using \eqref{2.51}-\eqref{2.53}, 
we obtain
\begin{eqnarray}
	V(t,s)-U(t,s) &=& \int_s^t V(t,r)K(r)U(r,s)d r\nonumber\\
	&=& \frac12\sum_{j=1}^d \int_s^t V(t,r)\big[ \G(r), \R_j\big(r,\partial_tP_j(r)\big) \big]U(r,s)d r.
	\label{48.0}
\end{eqnarray}
Now we use $V(t,r)\G(r)=-i\ve\partial_r V(t,r)-i\ve V(t,r)K(r)$ and $\G(r)U(r,s) = i\ve\partial_r U(r,s)$ to write
\begin{eqnarray}
	\lefteqn{
		\int_s^t V(t,r)\big[ \G(r), \R_j\big(r,\partial_tP_j(r)\big) \big]U(r,s)d r}\nonumber\\
	&=&-i\ve \int_s^t \Big[ \{ \partial_rV(t,r) + V(t,r)K(r)\} \R_j\big(r,\partial_tP_j(r)\big) U(r,s)\nonumber\\
	&& \qquad \ +V(t,r) \R_j\big(r, \partial_tP_j(r)\big) 
	\partial_rU(r,s)\Big] dr\nonumber\\
	&=& -i\ve\int_s^t \partial_r\big[V(t,r) \R_j\big(r,\partial_tP_j(r)\big) U(r,s)\big] d r\nonumber\\
	&& -i\ve\int_s^t V(t,r) \big[K(r) \R_j\big(r,\partial_tP_j(r)\big) -\partial_r\{ \R_j\big(r,\partial_tP_j(r)\big) \} \big] U(r,s)d r.\quad 
	\label{48.1}
\end{eqnarray}
We prove \eqref{estVU} below, which yields for all $0\le s\le t$
\begin{equation}
	\label{49.1}
	\big( \|U(t,s)\| +\|V(t,s)\|\big)\le C_0e^{\tilde C_0(t-s)}
\end{equation}
for constants $C_0, \tilde C_0$ independent of $\ve$ and $\lambda$. 
Next, we observe that the identity $\partial_tP_j(t)=(\partial_tP_j(t))P_j(t)+P_j(t)\partial_tP_j(t)$ allows us to simplify \eqref{2.53} with $X=\partial_tP_j(r))$ to
\begin{align}\label{simR}
    \R_j(r,\partial_tP_j(r))=\sum_{\ell=1,\ldots,d,\, \ell\neq j} \frac{\partial_tP_j(t)P_\ell(t) + P_\ell(t) \partial_tP_j(t)}{\widetilde\alpha_\ell(t)-\widetilde\alpha_j(t)},
\end{align}
so that we estimate 
\begin{equation}
\|\R_j(r,\partial_tP_j(r))\| \le \frac{2  d\, \|\partial_tP_j(r)\| \max_{1\le j\le d} \|P_j\|_\infty}{\min_{j,k\, :\, j\neq k}\,  \min_{0\le t\le 1} |\widetilde\alpha_j(t)-\widetilde\alpha_k(t)|}\le \frac{8  d}{\Delta_0} \, \|\partial_tP_j(r)\|. 
\label{2.63-1}
\end{equation}
To arrive at the second estimate in \eqref{2.63-1}, we
 use $\|P_j\|_\infty\le 2$ (see Lemma \ref{lambdasmall}(c)) and we estimate the denominator in terms of $\Delta_0$, \eqref{gap}, see also \eqref{1.43}: 
\begin{eqnarray}
|\widetilde\alpha_j(t)-\widetilde\alpha_k(t)| &=& \big|\alpha_j(t)-\alpha_k(t)+\lambda^2(\alpha'_j(t,\ve)-\alpha'_k(t,\ve))\big|\nonumber\\
&\ge& \big|\alpha_j(t)-\alpha_k(t)\big|-\lambda^2 |(\alpha'_j(t,\ve)-\alpha'_k(t,\ve))|\nonumber\\
&\ge& \Delta_0 -2\lambda^2 \|v\|_\infty^2\|\gamma\|_{L^1}\nonumber\\
&\ge& \Delta_0/2,
\end{eqnarray}
where we have used the condition \eqref{lambdasmall} in the last step. We now estimate the first term on the right side of \eqref{48.1} by
\begin{eqnarray}
    \ve \, \big\|\int_s^t\partial_r\big[V(r,t) \R_j\big(r,\partial_tP_j(r)\big) U(r,s)\big] d r \big\| &\le& \ve \|V(r,t) \R_j\big(r,\partial_tP_j(r)\big) U(r,s)\|\,\big|_{r=s}^{r=t}\nonumber\\
    &\le& \ve \frac{16 C_0d}{\Delta_0}\max_{r=s,t}  \|\partial_tP_j(r)\|e^{\tilde C_0(t-s)}\nonumber\\
&\le& C (\ve +\lambda^2)e^{\tilde C_0(t-s)},
    \label{2.66}
\end{eqnarray}
for a constant $C$ independent of $\vl$ and $0\le s\le t\le 1$. In the last step, we have made use of the estimate \eqref{1.34} on the derivative of the projection.

Next we estimate the second term on the right side of \eqref{48.1}. We start with  
\begin{eqnarray}
\ve\, \big\| \int_s^t V(t,r) K(r) \R_j\big(r,\partial_tP_j(r)\big)U(r,s)dr\big\| &\le& \ve C_0^2e^{\tilde C_0(t-s)}\int_s^t \|K(r)\|\, \|\R_j(r,\partial_tP_j(r))\| dr \nonumber\\
&\le& \ve \frac{8d^2 C_0^2}{\Delta_0} e^{\tilde C_0(t-s)} \max_j \int_s^t \|\partial_t P_j(r)\|^2 dr,
\label{2.66-1}
\end{eqnarray}
where we have taken into account \eqref{2.63-1} and that $\|K(r)\|\le 2d \max_j\|\partial_tP_j(r)\|$ (see \eqref{1.38.1} and use $\|P_j\|\le 2$). Next, according to \eqref{1.34}, $\|\partial_tP_j(r)\|^2 \le C\big( 1 + \frac{\lambda^2}{\ve}|\gamma(r/\ve)|+\frac{\lambda^4}{\ve^2}|\gamma(r/\ve)|^2\big)$, so
\begin{equation*}
\int_s^t \|\partial_tP_j(r)\|^2dr \le C\Big( t-s +\lambda^2\|\gamma\|_{L^1}+\frac{\lambda^4}{\ve}\|\gamma^2\|_{L^1} \Big).
\end{equation*}
Combining this bound with \eqref{2.66-1} yields
\begin{equation}
\ve\, \big\| \int_s^t V(t,r) K(r) \R_j\big(r,\partial_tP_j(r)\big)U(r,s)dr \big\| \le Ce^{\tilde C_0(t-s)}\big( \ve(t-s)+\ve\lambda^2+\lambda^4).
\label{2.60}
\end{equation}

Next we deal with the other term within the second integral on the right side of \eqref{48.1}, 
\begin{equation}
\ve \, \big\|\int_s^t V(t,r)\partial_r\{ \R_j\big(r,\partial_tP_j(r)\big) \}  U(r,s)d r\big\|\le \ve C_0^2e^{\tilde C_0(t-s)}\int_s^t  \|\partial_r\{\R_j(r,\partial_tP_j(r))\}\|dr.
\label{2.68}
\end{equation}
By using the quotient rule for derivatives in \eqref{simR}, 
\begin{multline}
\|\partial_r\{\R_j(r,\partial_tP_j(r))\}\| \le \frac{2d}{\Delta_0^2}\Big[\big(  \max_j\|\partial_tP_j(r)\|^2 \\
+\max_j\|P_j\|_\infty\max_j\|\partial^2_tP_j(r)\| \big)\max_{\ell, j} |\widetilde\alpha_\ell-\widetilde\alpha_j|_\infty\\
+\max_j\|P_j\|_\infty \max_j\|\partial_t P_j(r)\| \max_{\ell, j}|\partial_t(\widetilde \alpha_\ell - \widetilde\alpha_j)|_\infty \Big].
\label{2.62}
\end{multline}
Using the bounds $\|P_j\|_\infty\le 2$ (see before \eqref{1.34}) and $|\widetilde\alpha_\ell|_\infty\le |\alpha_\ell|_\infty+\lambda^2\|v\|_\infty^2 \|\gamma\|_{L^1}$, we simplify the estimate \eqref{2.62} to 
\begin{multline}
\|\partial_r\{\R_j(r,\partial_tP_j(r))\}\| \le \frac{8d}{\Delta_0^2} \Big[(\max_j|\alpha_j|_\infty +\lambda^2\|v\|_\infty^2\|\gamma\|_{L^1})\big(\max_j\|\partial_tP_j(r)\|^2 +\max_j\|\partial_t^2P_j(r)\|\big) \\
+\max_j\|\partial_tP_j(r)\| \max_j|\partial_t\widetilde\alpha_j|_\infty \Big].
\label{2.63}
\end{multline}
We have 
\begin{eqnarray}
|\partial_t\widetilde\alpha_j(t)| &\le& |\partial_t\alpha_j(t)| +\lambda^2\big|\partial_t\big[|v_j(t)|^2 \int_0^{t/\ve} e^{ix\alpha_j(t)}\gamma(x)dx\big]\big|\nonumber\\
&\le& |\partial_t\alpha_j|_\infty +\lambda^2\big[ 2 \|v\|_\infty\|\partial_tv\|_\infty \|\gamma\|_{L^1} + \|v\|_\infty\frac1\ve|\gamma(t/\ve)|\big].
\label{2.64}
\end{eqnarray}
Combining \eqref{2.63} and \eqref{2.64} gives the bound
\begin{equation}
\|\partial_r\{\R_j(r,\partial_tP_j(r))\}\| \le 
C\max_j \big( \|\partial_tP_j(r)\|(1+\frac{\lambda^2}{\ve}|\gamma(r/\ve)|) +\|\partial_tP_j(r)\|^2+ \|\partial_t^2P_j(r)\| \big).
\label{2.71}
\end{equation}
The norms $\|\partial_tP_j(r)\|$ and $\|\partial_t^2 P_j(r)\|$ are estimated in \eqref{1.34} and \eqref{1.35.3} and so \eqref{2.71} becomes
\begin{equation}
\|\partial_r\{\R_j(r,\partial_tP_j(r))\}\| \le 
C\big( 1+\lambda^2 + \frac{\lambda^2}{\ve^2}(|\gamma(r/\ve)|(1+r) + |(\partial_t\gamma)(r/\ve)|) +\frac{\lambda^4}{\ve^2}|\gamma(r/\ve)|^2 \big). 
\label{2.72}
\end{equation}
This bound inserted into \eqref{2.68}, and using $\int_s^t|\gamma(r/\ve)|rdr\leq \ve^2\|r\gamma(r)\|_{L^1}$, yields
\begin{multline}
\ve \, \big\|\int_s^t V(t,r)\partial_r\{ \R_j\big(r,\partial_tP_j(r)\big) \}  U(r,s)d r\big\| \le Ce^{\tilde C_0(t-s)}\big( \ve(t-s)(1+\lambda^2)+\ve \lambda^2\|r\gamma(r)\|_{L^1} \\\
 + \lambda^2( \|\gamma\|_{L^1}+\|\partial_t\gamma\|_{L^1}) +\lambda^4\|\gamma^2\|_{L^1}\big).
\label{2.74}
\end{multline}
Putting together the relations \eqref{48.0}, \eqref{48.1}, \eqref{2.66}, \eqref{2.60} and \eqref{2.74} we are rewarded with the bound
\begin{equation}
\|V(t,s)-U(t,s)\| \le Ce^{\tilde C(t-s)} ( \ve +\lambda^2 ), 
\label{2.75}
\end{equation}
where $\tilde C > \tilde C_0$, which holds for $\lambda$ satisfying \eqref{lambdasmall}, and all $0< \ve \leq 1$, $0\leq s\leq t$
as given in \eqref{42}.

Finally we need to check the validity of \eqref{estVU}, point 2) of Proposition \ref{prop4}, which in term implies \eqref{49.1}. To bound $V_\vl(t,s)$ we use the relation \eqref{41}. First we see that
\begin{equation}
	\label{1.61}
	\|W_\vl(t,s)\|\le e^{\int_s^t \|K_\vl(u)\|du},
\end{equation}
which follows from the series expansion based on  \eqref{38}. Now by Lemma \ref{lem1}, point (c), 
\begin{equation}
	\| K_\vl(t) \|  \le \sum_{j=1}^d  \| \partial_tP_j(t,\vl)P_j(t,\vl)\|
	\le 2d \max_j \|\partial_t P_j(t,\vl)\|.
\end{equation}
Then by \eqref{1.34},
\begin{equation}
	\int_s^t \|K_\vl(u)\|du \le  (t-s) C_1  + \|w\|_\infty^2 \frac{\lambda^2}{\ve}\int_s^t  |\gamma(u/\ve)|du,
	\label{1.63}
\end{equation}
with
\begin{equation}
	\label{1.53}
	C_1=\frac{8}{\Delta_0} \Big[ \|\partial_tA\|_\infty + \lambda^2\big( 2\|\partial_t w\|_\infty\|w\|_\infty \|\gamma\|_{L^1}+\|v\|_\infty^2 \|t\gamma(t)\|_{L^1}\, \|\partial_tA\|_\infty \big)\Big].
\end{equation}
As $\frac1\ve\int_s^t|\gamma(u/\ve)|du 
\le \|\gamma\|_{L^1}$, we conclude from \eqref{1.61} that for all $0\le s\le t<\infty$,
\begin{equation}
	\label{1.54.1}
	\|W_\vl(t,s)\| \le e^{(t-s) C_1+\lambda^2\|w\|_\infty^2\|\gamma\|_{L^1}}.
\end{equation}
This upper bound is uniform in $\ve>0$. The next step in our quest to control $\|V_\vl(t,s)\|$ is an upper bound on the phase term  $\Psi_\vl(t,s)$ in \eqref{41}. As $\|P_j(s,\vl)\|\le 2$ (Lemma \ref{lem1}(c)),
\begin{equation}
	\label{1.55.2}
	\|\Psi_\vl(t,s)\| \le 2d\max_j e^{\frac1\ve\int_s^t{\rm Im}\, [\alpha_j(u)+\lambda^2\alpha'_j(u,\ve)]du} = 2d \max_j e^{\frac{\lambda^2}{\ve} \int_s^t{\rm Im}\, \alpha'_j(u,\ve) du}.
\end{equation}
From \eqref{1.43} and the property $\gamma(-x)=\overline\gamma(x)$ (see \eqref{25.1}), we obtain
\begin{equation}
	{\rm Im}\, 	\alpha_j'(t,\ve) = - |v_j(t)|^2{\rm Re\,} \int_0^{t/\ve} e^{i x\alpha_j(t)}\gamma(x) dx = - \frac12|v_j(t)|^2\int_{-t/\ve}^{t/\ve} e^{i x\alpha_j(t)}\gamma(x) dx.
	\label{1.55.1}
\end{equation}
For any $0\le u_0<t-s$ we have
\begin{eqnarray}
	\int_s^t {\rm Im} \alpha_j'(u,\ve)du &=& \int_0^{t-s} {\rm Im} \alpha_j'(u+s,\ve)du \nonumber\\
	&=&\int_0^{u_0} {\rm Im} \alpha_j'(u+s,\ve)du +\int_{u_0}^{t-s} {\rm Im} \alpha_j'(u+s,\ve)du. 
	\label{1.59}
\end{eqnarray}
As $|\alpha'_j(t,\ve)|\le \|v_j\|^2_\infty\|\gamma\|_{L^1}$ (see \eqref{1.43}), the first integral on the right side of \eqref{1.59} is bounded above by $u_0\|v_j\|^2_\infty\|\gamma\|_{L^1}$. The second integral is
\begin{eqnarray}
	\lefteqn{ 
		-\frac12 \int_{u_0}^{t-s}  |v_j(u+s) |^2 \Big( \int_{-(u+s)/\ve}^{(u+s)/\ve} e^{i x\alpha_j(u+s)}\gamma(x) dx \Big)  du}\label{1.60}\\
	&=& -\frac12 \int_{u_0}^{t-s}  |v_j(u+s) |^2 \Big(\sqrt{2\pi} \widehat\gamma\big(\alpha_j(u+s) \big) -  \int_{|x|\ge (u+s)/\ve} e^{i x\alpha_j(u+s)}\gamma(x) dx \Big) du, \nonumber
\end{eqnarray}
where
\begin{equation}
    \widehat \gamma(\alpha) = \frac{1}{\sqrt{2\pi}} \int_\rx e^{ix\alpha}\gamma(x)dx\geq 0.
\end{equation}
Using the bound \eqref{1.59.1} we estimate
\begin{eqnarray}
	\lefteqn{
		\Big| \frac12 \int_{u_0}^{t-s}  |v_j(u+s) |^2  \Big(\int_{|x|\ge (u+s)/\ve} e^{i x\alpha_j(u+s)}\gamma(x) dx \Big) du\Big|} \nonumber \\
	&&\le\frac{C_\gamma\|v_j\|_\infty^2}{m-1}  \int_{u_0}^{t-s} \frac{du}{(1+\frac{u+s}{\ve})^{m-1}}\le \frac{C_\gamma\|v_j\|_\infty^2}{(m-1)(m-2)}  \frac{\ve}{(1+\frac{u_0}{\ve})^{m-2}}.
	\label{1.62}
\end{eqnarray}
Since $\widehat\gamma(x)\ge 0$, we obtain from \eqref{1.59}, \eqref{1.60} and \eqref{1.62}
\begin{equation}
	\int_s^t {\rm Im} \alpha_j'(u,\ve)du \le u_0 \|v_j\|_\infty \|\gamma\|_{L^1} + \frac{C_\gamma\|v_j\|_\infty^2}{(m-1)(m-2)}  \frac{\ve}{(1+\frac{u_0}{\ve})^{m-2}}.
\end{equation}
Upon choosing  $u_0=\ve$ we get for $0\le s\le t<\infty$,
\begin{equation}
\label{2.67}
	e^{\frac{\lambda^2}{\ve} \int_s^t{\rm Im}\, \alpha'_j(u,\ve) du} \le e^{\lambda^2[\|v_j\|^2_\infty \|\gamma\|_{L^1} +\frac{C_\gamma\|v_j\|_\infty^2}{(m-1)(m-2){2^{m-2}}}]}.
\end{equation}
Combining this bound with \eqref{41}, \eqref{1.54.1} and \eqref{1.55.2} we arrive at 
\begin{equation}
\|	V_\vl(t,s)\| \le 2d  e^{(t-s) C_1+\lambda^2 [(\|w\|^2_\infty+ \|v\|^2_\infty) \|\gamma\|_{L^1} +\frac{C_\gamma\|v\|_\infty^2}{(m-1)(m-2){2^{m-2}}}]},
	\label{1.63.1}
\end{equation}
where $C_1$ and $C_\gamma$ are from \eqref{1.53}, \eqref{1.59.1}. This shows the bound on $V_\vl$ in  \eqref{estVU}.

\bigskip
Next, we bound $\|U_\vl(t,s)\|\equiv \|U(t,s)\|$. We use the first equality in \eqref{48.0} and iterate it,
\begin{eqnarray}
	U(t,s) &=& V(t,s)+ \sum_{n\ge 1} (-1)^n\int_s^t dr_1\int_s^{r_1} d r_2 \cdots\int_s^{r_{n-1}}d r_n V(t,r_1)K(r_1)  V(r_1,r_2) K(r_2)\cdots\nonumber\\
	&&  \cdots V(r_{n-1},r_n)K(r_n)V(r_n,s).
	\label{2.90}
\end{eqnarray}
The bound on $V_\vl(t,s)$ in \eqref{estVU} implies
$$
\|V(t,r_1)K(r_1)\cdots K(r_n)V(r_n,s)\| \le (2d)^{n+1} e^{(t-s)C_1} e^{(n+1)\lambda^2 C_2}\|K(r_1)\|\cdots \|K(r_n)\|,
$$
from which we obtain, by \eqref{2.90},
\begin{multline}
\|U(t,s)\|\le 2d\, e^{(t-s)C_1}e^{\lambda^2 C_2}\sum_{n\ge 0}\frac{1}{n!}\big[2d e^{\lambda^2C_2}\int_s^t \|K(r)\|dr \big]^n\\
=2d\, e^{(t-s)C_1}e^{\lambda^2 C_2} \exp \big[2d e^{\lambda^2C_2}\int_s^t \|K(r)\|dr \big].
\label{2.91}
\end{multline}
Now as in \eqref{1.63} and following that bound, $\int_s^t\|K(r)\|dr \le (t-s)C_1+\lambda^2\|w\|^2_\infty \|\gamma\|_{L^1}$, so that \eqref{2.91} (upon possibly relabeling constants) indeed implies the bound on $U_\vl(t,s)$ in \eqref{estVU}. 
This completes the proof of Proposition \ref{prop4}.\qed

\bigskip
{\bf Acknowledgements.\ } M.M. was supported by a {\em Discovery Grant} from NSERC, the {\em National Sciences and Engineering Research Council of Canada}, as well as a one-month visiting position at the {Centre de Physique Th\'eorique Grenoble-Alpes} (CPTGA).
A.J. was partially supported by the CNRS program PICS (DEASQO),
and by the ANR grant NONSTOPS (ANR-17-CE40- 0006-01).


\begin{thebibliography}{}

\bibitem[AJPP]{AJPP}  Aschbacher, W., Jaksic, V., Pautrat, Y., Pillet, C.-A.: Topics in quantum statistical mechanics. In Open Quantum Systems III, 
{\it Lecture Notes in Mathematics}, {\bf 1882}, (2006)

\bibitem[AL]{AL}
R. Alicki, K. Lendi, Quantum Dynamical Semigroups and Applications, in: Lect. Notes Phys., vol. {\bf 717}, Springer Verlag,
2007


\bibitem[AE]{AE} Avron, J.E., Elgart,A., Adiabatic theorem without a gap condition, {\it Commun. Math. Phys.}, {203} (1999), p. 445--463. 

\bibitem[AFGG1]{AFGG} Avron, J.E.,  Fraas, M., Graf, G.M., Grech, P., Adiabatic theorems for generators of contracting evolutions, {\it Commun. Math. Phys.}, {314} (2012), p. 163--191.

\bibitem[AFGG2]{AFGG2}    Avron, J.E., Fraas, M., Graf G.M., Grech, P., {\it Landau-Zener Tunneling for Dephasing Lindblad Evolutions}, Commun. Math. Phys. {\bf 305} (3), 633-639 (2011)

\bibitem[ASY]{ASY} Avron, J.E., Seiler,R., Yaffe, L.G., Adiabatic theorems and applications to the quantum Hall effect, {\it Commun. Math. Phys.}, {\bf 110} (1987), p. 33--49.

\bibitem[B]{B} Berry, M. V., Quantal phase factors accompanying adiabatic changes. {\it Proc. Royal Soc. Lond. Math. Phys. Sci.} {\bf 392}, (1984), 45–57.

\bibitem[BF]{BF} Born, M., Fock, V., Beweis des Adiabatensatzes. {\it Z. Phys.}, {51} (1928), p. 165--180.

\bibitem[CHL]{CHL} Chruscinski, D., Hesabi, S., Lonigro, D., 
On Markovianity and classicality in multilevel spin-boson models, 
{\it Preprint} (2022)

\bibitem[CL]{CL} Chruscinski, D., Lonigro, D., Excitation-damping quantum channels, {\it Preprint} (2022)

\bibitem[CJKN]{CJKN} Cornean, H. D., Jensen, A., Kn\"orr H. K., Nenciu, G., On the adiabatic theorem when eigenvalues dive into the continuum, {\it Rev. Math. Phys.} {\bf 30}, 1850011 (2018).

\bibitem[D1]{Davies} Davies, E. B.,  Dynamics of a multilevel Wigner‐Weisskopf atom {\it J. Math. Phys.}, 15, (1974) p. 2036-2041 

\bibitem[D2]{D2} Davies, E. B., Markovian Master Equations, {\it Commun. Math. Phys.}, {\bf 39} (1974), p. 91--110.

\bibitem[DF1]{DF1} Derezinski, J., Fruboes, Renormalization of the Freidrichs Hamiltonian,
{\it Reports on Mathematical Physics},
{\bf 50}, (2002), Pages 433-438

\bibitem[DF2]{DF} Derezinski, J., Fruboes, R.  Fermi Golden Rule and Open Quantum Systems, In Open Quantum Systems III, 
{\it Lecture Notes in Mathematics}, {\bf 1882}, (2006)

\bibitem[DK]{DK}Davidson, R., Kozak, J. J.,  On the Relaxation to Quantum Statistical Equilibrium of the Wigner Weisskopf Atom in a One Dimensional Radiation Field. I. A Study of Spontaneous Emission, {\it J. Math. Phys.}, {\bf 11}, 189 (1970)

\bibitem[DS]{DS} Davies, E. B., Spohn, H., Open Quantum Systems with Time-Dependent Hamiltonians and Their Linear Response, {\it J. Stat. Phys.}, {\bf 19} (1978), p. 511--523.
 

\bibitem[EN]{EN} Engel, K.-J., Nagel, R., One-parameter semigroups for linear evolution equations, Springer, 2000.


\bibitem[JKP]{JKP}Jaksic, V., Kritchevski, E., Pillet C.-A., Mathematical theory of the Wigner-Weisskopf atom, in Large Coulomb Systems, J. Derezinski, H. Siedentop Eds., {\it Lecture Notes in Physics} {\bf 695}, 145-215, 2006

\bibitem[J1]{J2}  Joye, A.,  General Adiabatic Evolution with a Gap Condition,  {\it Commun. Math. Phys.}, {275} (2007), p. 139--162.


\bibitem[J2]{J} Joye, A., Adiabatic Lindbladian Evolution with Small Dissipators, {\it Commun. Math. Phys.}, {\bf 391}, (2022), 223-267.

\bibitem[JMS]{JMS} Joye, A., Merkli, M., Spehner, D., Adiabatic transitions in a two-level system coupled to a free Boson reservoir, 
{\it Ann. H. Poincar\'e}, {\bf 21}, (2020), p. 3157-3199, 


\bibitem[K1]{K1} Kato,  T., On the Adiabatic Theorem of Quantum Mechanics, {\it J. Phys. Soc. Japan}, {5} (1950), p. 435--439.

\bibitem[K2]{K2} Kato,  T., Perturbation Theory for Linear Operators (Springer-Verlag Berlin Heidelberg New York 1980).


\bibitem[Kr]{Kr}
{Krein, S.G.}, Linear Differential Equations in Banach Space, {\it Translations of Mathematical Monographs, vol. 29}, AMS, 1971. 

\bibitem[Ma]{Ma} Martin, Ph. A., Mod\`eles en m\'ecanique statistique des processus irr\'eversibles, {\it Lecture Notes in Physics}, {\bf 103}, Springer, 1979

\bibitem[M1]{M1}
M. Merkli: {\em Quantum Markovian master equations: Resonance theory shows validity for all time scales}, Ann. Phys. {\bf 412}, 16799 (2020)  

\bibitem[M2]{M2}
M. Merkli: {\em Dynamics of Open Quantum Systems I, Oscillation and Decay}, Quantum {\bf 6}, 615 (2022)    

\bibitem[M3]{M3}
M. Merkli: {\em Dynamics of Open Quantum Systems II, Markovian Approximation}, Quantum {\bf 6}, 616 (2022)    
   





\bibitem[N1]{N1} Nenciu,  G., On the adiabatic theorem of quantum mechanics, {\it J. Phys. A}, Math. Gen., {13} (1980), p. 15--18.

\bibitem[N2]{N2} Nenciu, G., Existence of the spontaneous pair creation in the external
field approximation of Q.E.D. {\it Commun. Math. Phys.} {\bf 109}(2), 303–312 (1987)

\bibitem[NR]{NR} Nenciu, G.,  Rasche, G., On the adiabatic theorem for nonself-adjoint Hamiltonians, {\it J. Phys. A} {\bf 25}, (1992), 5741-5751.

\bibitem[Ne]{Ne}
 A.I. Nesterov, G.P. Berman, M. Merkli, A. Saxena: {\em Modeling of noise-assisted quantum transfer between donor and acceptor with finite bandwidths}, J. Phys. A: Math. Theor. {\bf 52}, 435601 (2019)

\bibitem[PD]{PD} Pickl, P., D\"urr, D., On Adiabatic Pair Creation. {\it Commun. Math. Phys.} {\bf 282}(1),
161–198 (2008)


\bibitem[Sc]{Sch} Schmid, J.,  Adiabatic theorems with and without spectral gap condition for non- semisimple spectral
values. In: Exner, P., K\"onig, W., Neidhardt, H. (eds.) Mathematical Results in Quantum Mechanics: Proceedings of the QMath12 Conference. World Scientific Publishing, Singapore, 2014.



\bibitem[Tr]{Tr}
A.S. Trushechkin, M. Merkli, J.D. Cresser, J. Anders: {\em Open quantum system dynamics and the mean force Gibbs state}, AVS Quantum Sci. {\bf 4}, 012301 (2022)    


\bibitem[Te]{T}  Teufel, S., A note on the adiabatic theorem without gap condition, {\it Lett. Math. Phys.},  {58} (2001), p.~261--266.

\bibitem[WW]{WW} Weisskopf, V., Wigner, E., Berechnung der naturlichen Linienbreite auf Grund der
Diracschen Lichttheorie. {\it Zeitschrift fur Physik} {\bf 63}, 54-73 (1930).




\end{thebibliography}
\end{document}